\documentclass[rmp,aps]{revtex4-1}
\usepackage[utf8]{inputenc}
\usepackage{xcolor}
\usepackage{ulem}
\usepackage{mathtools}
\usepackage{graphicx}
\usepackage{comment}

\newcommand{\be}{\begin{equation}}
\newcommand{\ee}{\end{equation}}

\def\m{{m}}
\def\M{{M}}
\def\p{{p}}
\def\P{{P}}

\graphicspath{{images/}}

\begin{document}

\title{Gene expression in growing cells: A biophysical primer{\footnote{To be submitted to {\it Reviews of Modern Physics}.}}}


\author{Ido Golding}

\affiliation{Department of Physics, University of Illinois at Urbana-Champaign, Urbana, IL, USA}

\affiliation{Department of Microbiology, University of Illinois at Urbana-Champaign, Urbana, IL, USA}

\author{Ariel Amir} 

\affiliation{John A. Paulson School of Engineering and Applied Sciences, Harvard University, Cambridge, MA, USA} 

\affiliation{Department of Physics of Complex Systems, Weizmann Institute of Science, Rehovot, Israel}

\date{\today}

\begin{abstract}

Cell growth and gene expression, two essential elements of all living systems, have long been the focus of biophysical interrogation. Advances in experimental single-cell methods have invigorated theoretical studies into these processes. However, until recently, there was little dialog between the two areas of study. In particular, most theoretical models for gene regulation assumed gene activity to be oblivious to the progression of the cell cycle between birth and division. But, in fact, there are numerous ways in which the periodic character of all cellular observables can modulate gene expression. The molecular factors required for transcription and translation—RNA polymerase, transcription factors, ribosomes—increase in number during the cell cycle, but are also diluted due to the continuous increase in cell volume. The replication of the genome changes the dosage of those same cellular players but also provides competing targets for regulatory binding. Finally, cell division reduces their number again, and so forth. Stochasticity is inherent to all these biological processes, manifested in fluctuations in the synthesis and degradation of new cellular components as well as the random partitioning of molecules at each cell division event. The notion of gene expression as stationary is thus hard to justify.
In this review, we survey the emerging paradigm of cell-cycle regulated gene expression, with an emphasis on the global expression patterns rather than gene-specific regulation. We discuss recent experimental reports where cell growth and gene expression were simultaneously measured in individual cells, providing first glimpses into the coupling between the two, and motivating several questions. How do the levels of gene expression products -- mRNA and protein -- scale with the cell volume and cell-cycle progression? What are the molecular origins of the observed scaling laws, and when do they break down to yield non-canonical behavior? What are the consequences of cell-cycle dependence for the heterogeneity (“noise”) in gene expression within a cell population? While the experimental findings, not surprisingly, differ among genes, organisms, and environmental conditions, several theoretical models have emerged that attempt to reconcile these differences and form a unifying framework for understanding gene expression in growing cells.
    
\end{abstract}

\maketitle

\tableofcontents

\subsection*{Prolog: Simple physical models of cellular processes} \label{prolog}


In this Colloquium, we discuss biophysical models for the process of gene expression, and how this process is coupled to the progression of the cell cycle (these terms will be elaborated below). Before embarking on this discussion, we should say a few words about the general nature of physical models for cellular processes. This preamble is required, we believe, because those models are somewhat different, in terms of how they are constructed and used, from physical models of inanimate matter, with which some readers are perhaps more familiar. These differences, in turn, reflect the vast gap in knowledge and experimental amenability between the physics of living and non-living systems.

One category of biophysical models aims for a molecular, or even atomic, level of description. Such models have been extremely successful in elucidating the function of biological molecules \cite{nelson2003biological, dill2011molecular}. However, even with advances in computational power, these models are limited to depicting just one or several molecules, over very short time scales (less than a millisecond). This makes the approach inadequate for capturing all but the simplest processes in the living cell, since these processes—such as the expression of genetic information, discussed below—typically involve numerous molecular players and take place over minutes and hours. Making a molecularly detailed model of these processes is thus currently impractical\footnote{There have been recent attempts at constructing models for cellular function, which explicitly consider thousands of molecular players \cite{karr2012whole, thornburg2022fundamental}. But while these models use statistical inference approaches to address the challenge of parameterization, they leave unresolved the problem of additional unrecognized regulatory interactions in the system.}. In truth, even if such ``full" cellular models were possible, to many physicists such a model would be unsatisfactory in its complexity -- Jorge Luis Borges's story comes to mind, of ``a Map of the Empire whose size was that of the Empire" \cite{borges1999collected}. For these reasons, cellular processes are typically conceptualized using simplified theoretical models, where a small number of molecules and interactions are considered explicitly, while many others are ignored or coarse-grained into other observables \cite{bintu2005transcriptional,  bialek2012biophysics, amir2018learning}. One need not be apologetic about the use of such phenomenological models. They tend to be more robust to the model details and lend themselves better to analytical approaches, which in turn can provide deeper insights into the physical principles at play \cite{phillips2005modeling,amir2020thinking}. 

In some ways, this approach follows the traditional physics attitude as applied, e.g., to the description of fluids or elastic materials, where countless microscopic constituents are left out \cite{taylor2005classical}. This omission is both a necessity—the positions and interactions of all atoms in the material are unknowable to us—and a choice, since it allows us to obtain a simple yet predictive depiction of the system. Biophysical models, too, reflect the combined constraints of ignorance and parsimony. However, much more than in the physics of non-living matter, the level of abstraction in biophysical models—what molecules, processes and interactions are included—reflects our ignorance of the underlying details. This ignorance, in terms of what is known or is even experimentally knowable, is overwhelming to a degree that is unfamiliar, and would perhaps be unacceptable, to modelers of nonliving systems. 

Even for the best characterized cellular processes, such as the regulation of gene expression in bacteria, a major focus of this Colloquium, what we currently have is a partial list of the molecular players involved and the interactions between them, but little or no knowledge of the biophysical parameters characterizing these interactions, such as the rates of diffusion, binding, and assembly into molecular complexes. Similarly, what we can experimentally measure is, too, highly limited in terms of number of molecular species simultaneously detected (typically, only a few), the precision (typically, relative rather than absolute levels, averaged over many individual cells), and the temporal resolution, which typically under-samples much of the relevant kinetics.     
 
The simplicity of biophysical models often reflects this limited knowledge regarding the systems under study, rather than an informed choice of which features to include and which ones to leave out.  In other words, the ``coarse-graining” process is driven by the need to remain anchored in known facts and make model predictions experimentally testable. \footnote{Tellingly, even the term ``coarse-graining" possesses a different meaning here compared with statistical or condensed matter physics. Whereas in those contexts it refers to a well-defined mathematical procedure, in the physics of living systems it is used more loosely, indicating an attempt to describe the system at some lower resolution, without accounting for all of the underlying processes - even if there is no rigorous procedure for finding the appropriate level of description or the number of parameters necessary.}. Thus, while there are select examples where a simple biophysical model may be argued to reflect an underlying simplicity of behavior, a-la Occam's razor (that of bacterial ``growth laws" \cite{scott2010interdependence}, relating ribosome levels to growth rates, is discussed later), in most instances model simplicity instead implies that we are ignorant of many details, which are swept under the proverbial ``Occam's rug" \cite{BRENNER1997R202, golding2011decision}. In addition to the many ``known unknowns", for example, the rate constants of the regulatory interaction under study, even more worrisome are the ``unknown unknowns", e.g., the presence of additional unrecognized interactions in the system. Making models more elaborate may make them appear more realistic, but typically only achieves the opposite, since the added details are inevitably less grounded in knowledge. Model elaboration can only be justified as a means to explore specific hypotheses that can be experimentally tested.   
 
In this Colloquium, we will describe how the constraints discussed above have driven the development of models describing gene expression and its coupling to cell growth (with the accompanying changes in the amount of molecules driving gene expression--DNA, RNA polymerase, ribosomes). We will often focus on bacterial systems, where the knowledge infrastructure and the experimental tractability are significantly superior to the more complex eukaryotic and multicellular systems. But, as we have emphasized already, even for bacteria our ignorance is—by physics standards—overwhelming, and this ignorance has strong consequences for constructing theoretical models. 


\section{Models of gene expression}
\label{chapter-static}

\subsection{Gene expression models with constant rates} \label{model-const-rates}

In the process of gene expression, a segment of the cell’s genome (the gene) is transcribed repeatedly into a complementary, short lived messenger RNA (mRNA). Each mRNA molecule is then translated, again repeatedly, into the protein encoded by that gene \cite{Albe_2002_book}. Since each cell's identity, shape, and function are largely determined by which proteins it expresses, gene expression and its regulation can be seen as the prime mover in the living cell\footnote{The preceding statement, as always in biology, has important exceptions, such as the cell's use of post-translational protein modifications to encode additional information \cite{Albe_2002_book}. This subject is outside the scope of the current Colloquium.}.

In constructing a biophysical model for gene expression, we must first note that the production of even a single protein molecule involves thousands of stochastic molecular events. On the transcription side alone, these events include RNA Polymerase (RNAP) and transcription factors (TFs) searching the whole cell and genome to find the regulatory region of the gene (called {\it promoter}) and binding to it; changes in molecular conformation of RNAP that enable the initiation of transcription; and the basepair-by-basepair synthesis of mRNA by RNAP, until the gene termination site is reached. The synthesis of protein from mRNA, and the degradation of both mRNA and proteins, are likewise molecularly elaborate \cite{Albe_2002_book}. Moreover, these different molecular events are regulated by multiple cellular factors, and subject to feedback from downstream steps in the gene expression processes, in ways that are only partly understood \cite{berry2022mechanisms}. 

\begin{figure}[htp]
    \centering
    \includegraphics[width=8cm]{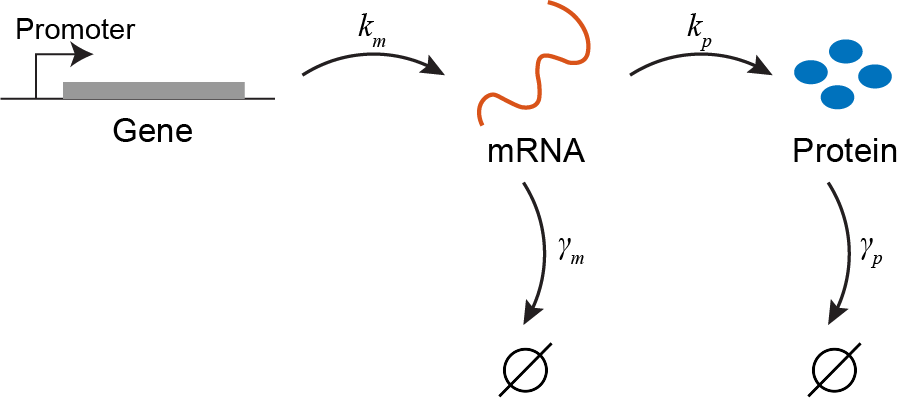}
    \caption{A minimal stochastic model for gene expression. }
    \label{fig:model1}
\end{figure}

Despite this complexity, gene regulation is often modeled using a mere four rates, corresponding to the production and degradation of mRNA (the copy number of which is denoted by $\M$) and proteins (whose copy number is denoted by $\P$) \cite{paulsson2005models, swain2002intrinsic, ozbudak2002regulation} {(Fig. \ref{fig:model1}}).  

This stochastic model can be succinctly summarized as:
\be \M \xrightarrow{k_m}  \M+1 , \label{model1}\ee
\be \M \xrightarrow{\gamma_m \M }  \M-1 , \label{model2}\ee
\be \P \xrightarrow{k_p \M }  \P+1 , \label{model3}\ee
\be \P \xrightarrow{\gamma_p \P}  \P-1. \label{model4}\ee
Here, $k_m$ denotes the transcription rate, $k_p$  the translation rate (per mRNA), and $\gamma_m$, $\gamma_p$ the degradation rates for mRNAs and proteins, respectively\footnote{A note on notation: Throughout the Colloquium, the distinction between molecule copy number and its concentration will be important. We will use $\M$, $\P$ to denote mRNA and protein copy numbers, respectively, and $\m$, $\p$ for their concentrations.}. 

As can be surmised from the earlier discussion, coarse graining the molecular complexity of gene expression into this simple standard model does not so much reflect informed choices as an intuitive attempt at parsimony, whose legitimacy depended on the limited resolution of experimental data available until about two decades ago. However, once experimental methods improved to allow measuring gene expression at finer resolution, the inadequacy of this simple model was revealed, leading to necessary modifications, as we shall discuss in Section \ref{model-two-state}.

But first, let us examine the model in some detail. We will first consider the \textit{ensemble means} of the observables, for which we may write down a set of readily solvable ODEs:
\be \frac{d\overline{\M}}{dt} = k_m - \gamma_m \overline{\M}.\label{ODE1}\ee
\be \frac{d\overline{\P}}{dt} = k_p \overline{\M} -  \gamma_p \overline{\P} .\label{ODE2}\ee

\begin{figure}[htp]
    \centering
    \includegraphics[width=8cm]{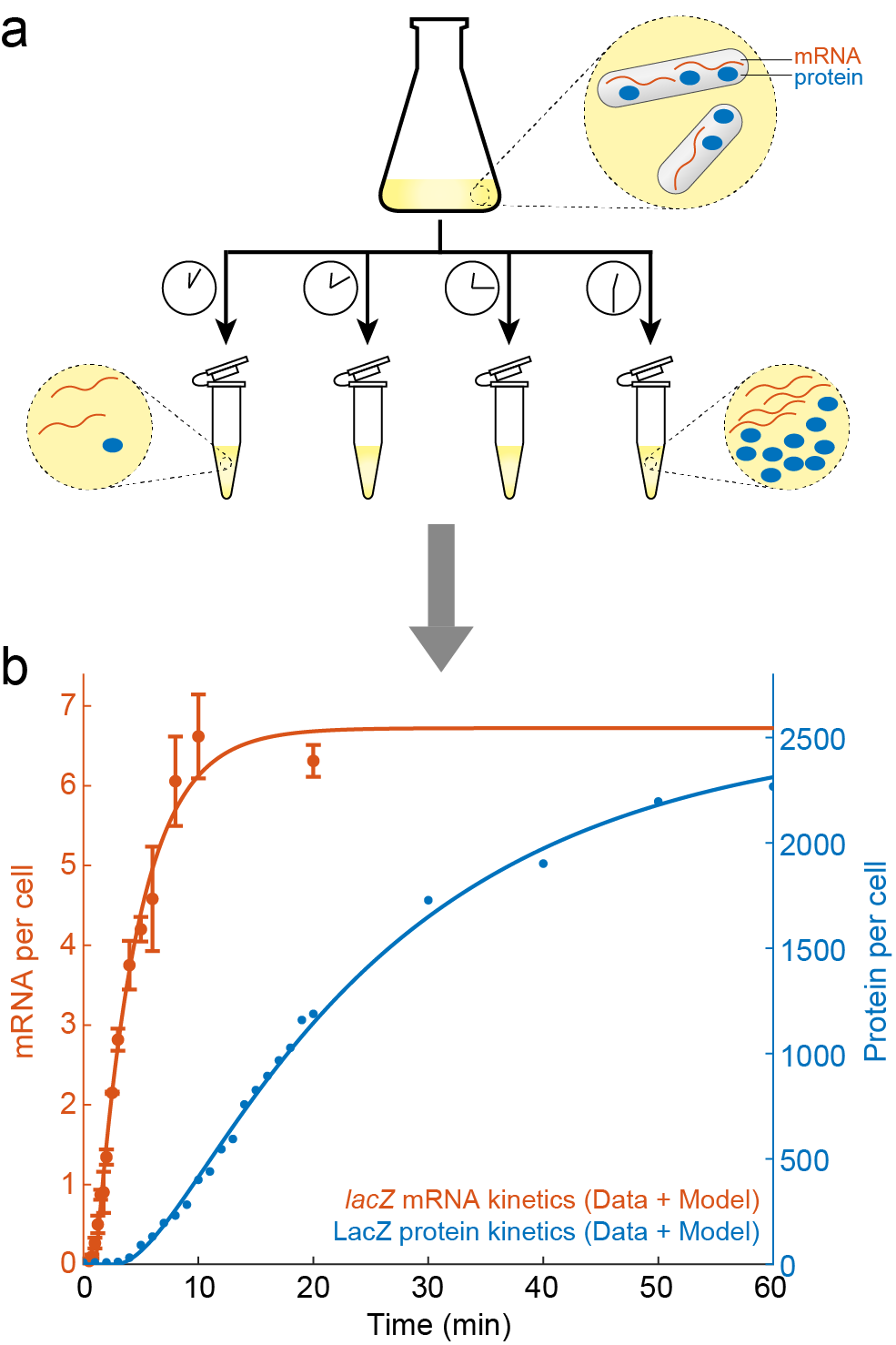}
    \caption{Bulk measurement of mRNA and protein levels. 
    (a) Millions of cells are grown in a flask, samples are taken at different times, and the cellular contents extracted from the cells. The amount of specific mRNA or protein in the sample can then be read by various means, e.g., biochemically amplifying the mRNA to a detectable amount, and assaying the enzymatic activity of the protein.
    (b) mRNA and protein kinetics during gene induction. {\it E. coli} cells were grown in glycerol media, and expression of the {\it lacZ} gene was induced by adding Isopropyl $\beta$-D-1-thiogalactopyranoside (IPTG). 
    mRNA data is from \cite{wang2019measuring}, protein data by Seunghyeon Kim and Sangjin Kim (unpublished; see \cite{kim2019long} for method). The experimental data is captured by Eqs. \ref{ODE1}-\ref{ODE2}. Model fitting by Tianyou Yao and Yuncong Geng.  
    }
    \label{fig:bulk-msrmnt}
\end{figure}

The ``ensemble” here can be interpreted as consisting of the individual cells within a population. Since traditional biochemical methods for measuring mRNA and protein levels are typically performed in bulk, using millions of cells to obtain a single reading {(Fig. \ref{fig:bulk-msrmnt})}, the ensemble average is the natural observable to be calculated. Performing these experimental measurements, one finds that Eqs. (\ref{ODE1})-(\ref{ODE2}) neatly capture mRNA and protein kinetics during gene induction, i.e., when the gene is turned ``on” {(Fig. \ref{fig:bulk-msrmnt})}. In other words, the standard model for gene expression appears to be consistent with the experimental data.

\begin{figure}[htp]
    \centering
    \includegraphics[width=8cm]{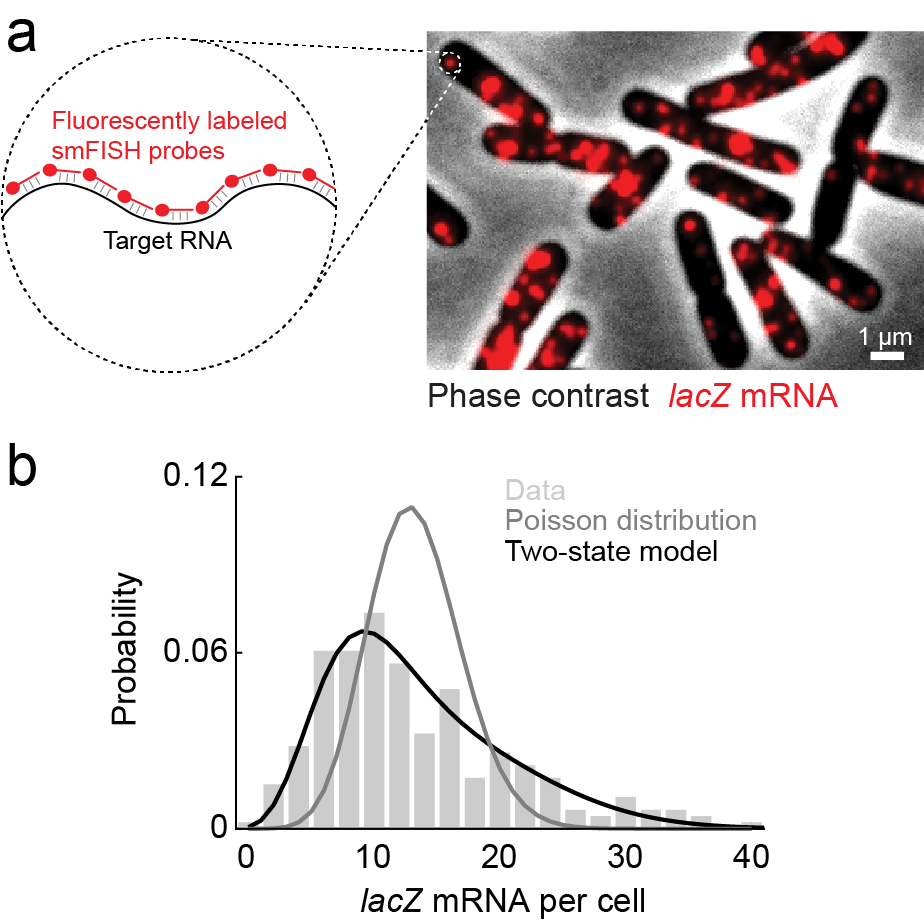}
    \caption{Single-cell measurement of mRNA copy-number.
   (a) Single-molecule fluorescence in situ hybridization (smFISH) is used to detect mRNA molecules in individual {\it E. coli} cells.
   (b) The measured distribution of mRNA copy number deviates from Poisson statistics but is well described by a two-state model.   
   Image and data are reproduced from \cite{wang-rnap-book} with permission from the Royal Society of Chemistry.}
    \label{fig:smfish}
\end{figure}

More recently, however, it has become possible to measure mRNA and protein numbers in an individual cell \cite{skinner2013measuring, cai2006stochastic, yu2006probing, taniguchi2010quantifying} {(Fig. \ref{fig:smfish})}, thus allowing us to go beyond the population mean and examine the copy number distribution. Characterizing the statistics -- rather than the mean alone -- of expression level is significant for two reasons. First, cell-to-cell differences in protein levels may result in variations in phenotype, such that genetically identical cells, within a uniform environment, diverge in their behavior. This cellular individuality plays a crucial role throughout biology, from the emergence of antibiotic persistence among bacteria, to cell differentiation in the early mammalian embryo, and numerous other examples \cite{balazsi2011cellular, eldar2010functional}. Thus, describing gene expression in individual cells is arguably more important than capturing it in the hypothetical ``average cell".

But there is also a second, biophysical reason to examine single-cell expression, and that is to provide stronger empirical challenge to the theoretical picture we presented 
in Fig. \ref{fig:model1}. To do so, we will interrogate the model further by considering the stochastic fluctuations associated with it, and derive theoretical results regarding the copy-number statistics, which can then be compared to experimental data. As we shortly see, this exercise will prove insightful.   

We begin by finding the steady-state distribution of mRNA copy number. To this end, we can write the master equation for the temporal dynamics of this distribution, $\mathcal{P}_M(n,t)$ (the probability to have precisely $n$ copies of mRNA as time $t$)\footnote{We will use $\mathcal{P}$ to denote both discrete and continuous probability distributions (i.e., probability density functions in the mathematics notation).}:
\be \frac{d\mathcal{P}_M(n,t)}{dt} = k_m \mathcal{P}_M(n-1,t) +  (n+1) \gamma_m \mathcal{P}_M(n+1,t) - (k_m + n \gamma_m) \mathcal{P}_M(n,t). \label{master_mrna} \ee
Here, the first and second terms on the RHS correspond, respectively, to the incoming fluxes from production of an mRNA when there were $n-1$ copies in the cell, or degradation of a molecule when there were $n+1$ copies in the cell. The last term on the RHS corresponds to degradation and production events that change $n$ copies in a cell to $n-1$ and $n+1$, respectively. 

At steady-state, $\frac{d\mathcal{P}_M(n,t)}{dt}=0$, and we can therefore write:
\be \gamma_m n \mathcal{P}_M(n) + k_m \mathcal{P}_M(n) = (n+1)\gamma_m  \mathcal{P}_M(n+1) + k_m \mathcal{P}_M(n-1). \label{mrna_dist}\ee 
What should we expect the solution for this equation to look like? If every mRNA molecule did not decay stochastically, but instead lived for precisely a time $1/\gamma_m$, then the number of mRNA molecules within the cell at a given time would equal the total number of molecules produced within a time-window $1/\gamma_m$ -- which is, of course, Poisson distributed with parameter $k_m/\gamma_m$. One can verify by direct substitution that this is also the exact solution to Eq. (\ref{mrna_dist}). In fact, if the initial mRNA copy number is Poisson distributed, it can be shown, by substitution into Eq. (\ref{master_mrna}), that the distribution will be Poissonian at \textit{all times}, albeit with a time-dependent parameter $\lambda(t)$ obeying the ODE:
\be \frac{d\lambda}{dt} = - (\lambda \gamma_m - k_m). \ee

Next, we turn to the steady-state distribution of \textit{protein} numbers, which is a little trickier to handle. Before delving into the equations, let us consider the stochastic kinetics of mRNA and protein numbers. Once an mRNA molecule is transcribed (a process occurring with rate $k_m$), proteins will start being produced at a rate $k_p$. This will happen until the mRNA is degraded, the timing of which is exponentially distributed (and is typically much shorter than protein lifetime, hence we can ignore protein degradation for the purpose of this calculation). We will refer to the event where multiple proteins are translated from a single mRNA copy as a \textit{burst}. Since protein production from each mRNA occurs at a constant rate, and since the mRNA lifetime distribution is exponential, the distribution $\nu(b)$ of the number of proteins $b$ produced in a single burst will be given by:
\be \nu(b) = \int_0^\infty \gamma_m e^{-t \gamma_m } Poiss(b, k_p t) dt,\ee 
with $Poiss(x,\lambda)$ the Poisson distribution with parameter $\lambda$. This integral can be readily evaluated, resulting in a geometric distribution for the protein copy number produced within a burst:
\be \nu(b) = \left(\frac{k_p}{k_p+\gamma_m}\right)^b \frac{\gamma_m}{k_p+\gamma_m}. \label{burst} \ee
The average burst size, $\bar{b}$, is found to be, as expected:
\be \bar{b} = \frac{k_p}{\gamma_m}. \label{burst_size}\ee

To proceed and find the protein copy number distribution, it will be convenient to work with a \textit{continuous} (i.e., Fokker-Planck) equation \cite{friedman2006} rather than a discrete one, as we did in the case of mRNA above. Clearly, for the existence of a stationary protein distribution we will need to consider a finite protein degradation rate (which was inconsequential for the previous calculation) -- otherwise proteins will continue to accumulate indefinitely. We will also assume that $\bar{b} \gg1$ such that the burst size distribution may be approximated as continuous. The continuous approach will be inaccurate when the protein level in a cell is low, but in practice these levels are often sufficiently high to make the results we will obtain a useful approximation (we will comment on the exact solution of the discrete equations shortly). The Fokker-Planck equation reads:
\be \frac{\partial \mathcal{P}}{\partial t} =\frac{\partial [\gamma_p x \mathcal{P}(x,t)]}{\partial x } + k_m \int_0^\infty \nu(b) \mathcal{P}(x -b ,t) db -k_m \mathcal{P}(x,t), \label{FP} \ee
where $x$ is the (now continuous) protein copy-number and the probability distribution $\nu(b)$ is given by Eq. (\ref{burst}). 
Equating the time derivative to zero leads us to an equation for the the steady state. One may verify by direct substitution that the (normalized) solution is given by the gamma distribution \cite{friedman2006}):
\be \mathcal{P}(x) = \frac{1}{\bar{b}^a \Gamma(a)} x^{a-1} e^{-x/\bar{b}}, \label{stat_solution}\ee
with $a = k_m/\gamma_p$, assumed to be a large number, and thus $\Gamma(a) \approx [a-1]!$, with $[]$ indicating the nearest integer.

In fact, this form may be intuited by noting that a sum of $n$ independent variables, each drawn from an exponential distribution, is gamma distributed with a shape parameter $n$ (see chapter 6 of \cite{amir2020thinking}). In our case, each exponentially distributed variable corresponds to a single burst of proteins. Since the protein lifetime is $1/\gamma_p$, we expect $n$ to equal the number of bursts in this time window, namely $n= k_m/\gamma_p=a$ -- which turns out to be the precise result\footnote{The fact that the sum of $n$ independent, exponentially-distributed variables is gamma distributed is helpful in evaluating the convolution which appears in Eq. (\ref{FP}), when verifying that Eq. (\ref{stat_solution}) is a stationary solution of Eq. (\ref{FP}).}. We note that the discrete case can also be solved, and in the limit of short mRNA lifetime yields a negative binomial distribution \cite{raj2006stochastic, shahrezaei2008analytical}. 

\subsection{Comparison to experimental data and the two-state model for gene expression} \label{model-two-state}
Now that we have found theoretical predictions for the distributions of mRNA and protein copy numbers, we turn to the experimental data \cite{golding2005real, so2011general}. As can be seen in {Fig. \ref{fig:smfish}}, the measured mRNA distribution is, alas, very poorly fit by the Poisson distribution we predicted above. Intriguingly, the mRNA data is found to be well-fitted by a gamma (or a negative binomial) distribution -- which was the (approximate) result we expected for the protein number distribution. What do we learn from this conundrum? 

\begin{figure}[htp]
    \centering
    \includegraphics[width=8cm]{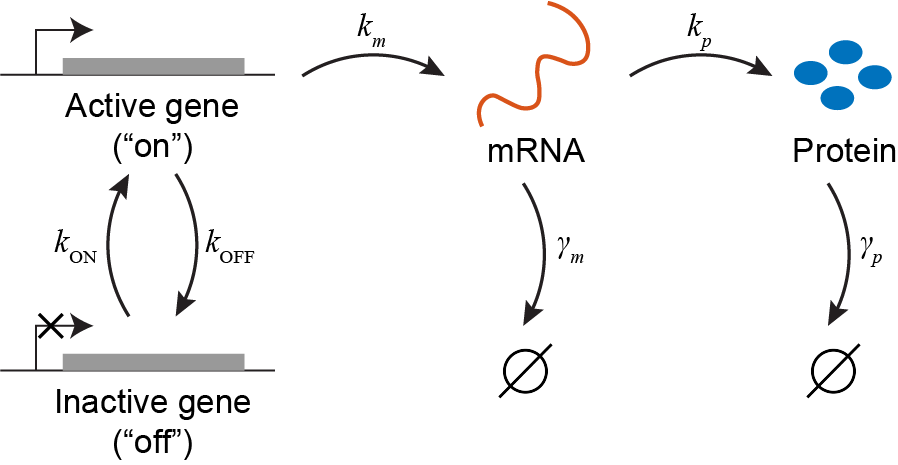}
    \caption{The two-state model for stochastic gene expression. }
    \label{fig:model2}
\end{figure}

The insight lies in realizing that the gamma distribution arose from a model where one molecular species follows a birth-death process (i.e., it is produced and decays at constant rates) and a second species is made at a rate proportional to the copy number of the first one. We may, effectively, obtain the same result for the mRNA distribution if we postulate that the gene from which mRNA is produced can be either ``on" or ``off", and that the switching between the two states occurs at constant rates -- see Fig. \ref{fig:model2}. This model, commonly referred to as the ``two-state" (or ``telegraph") model \cite{paulsson2005models}, will produce bursts of mRNA, that -- since the stochastic dynamics is formally identical to that of protein production in the simpler, one-state, model analyzed earlier -- will be exponentially distributed. A Fokker-Planck equation, analogous to Eq. (\ref{FP}), can then be set up for the steady-state mRNA copy number distribution, leading to the gamma distribution (or, if we treat mRNA numbers as discrete rather than continuous, a negative binomial distribution).

\begin{figure}[htp]
    \centering
    \includegraphics[width=8cm]{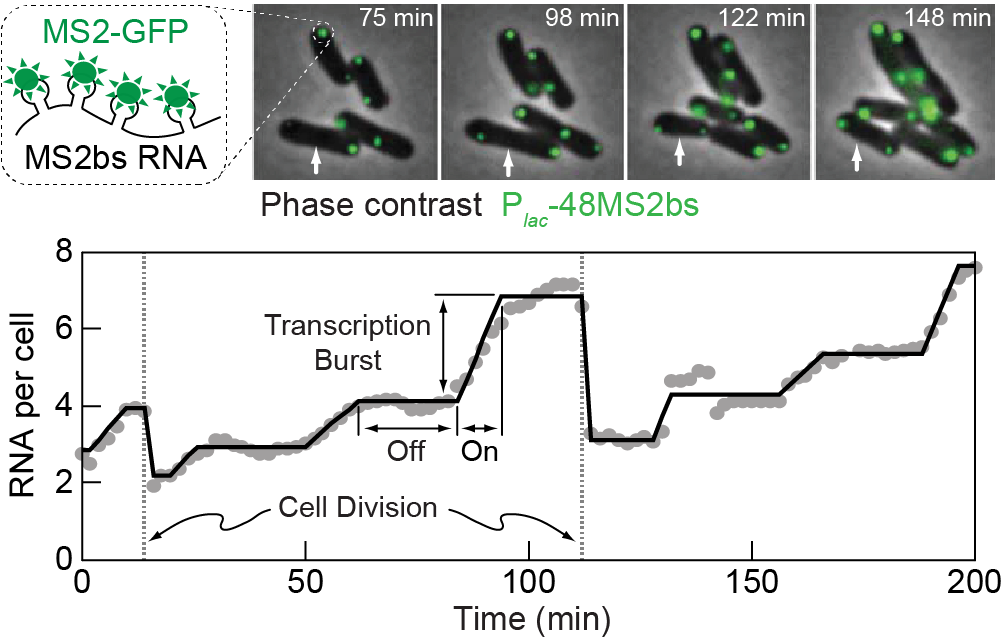}
    \caption{Following mRNA kinetics in live cells reveals transcription bursts. 
    Top, transcription is followed in real time by labeling mRNA with a genetically-encoded fluorescent protein (MS2-GFP). Bottom, the resulting time series from a single cell (marked above with a white arrow) exhibits bursts of transcription, such as the one highlighted at $\approx$ 90 minutes, consistent with the prediction of the two-state model. Data by Lok-Hang So and Ido Golding, reproduced from \cite{phillips2013physical} with permission of Taylor \& Francis through PLSclear. 
     }
    \label{fig:bursting}
\end{figure}

The two-state model for transcription is able to capture mRNA statistics both at steady-state and during gene induction, and was further validated by following the stochastic kinetics of mRNA production in live cells {(Fig. \ref{fig:bursting})}, which exhibits the exponentially distributed transcription ``bursts" predicted by the model \cite{golding2005real}. Note that for the \textit{ensemble-average} behavior, once we coarse-grain over the timescale of gene switching, the model reduces to the naive model we started with -- and thus the agreement with the bulk experiments is retained \cite{golding2005real}.  Beyond bacteria, the two-state model has been shown to reproduce mRNA statistics in higher organisms, from yeast to mammalian tissues \cite{sanchez2013genetic, skinner2016single}. Notwithstanding this success, the mechanistic basis of gene on/off switching is still debated. The mechanism likely varies between different genes and organisms, with roles proposed for transcription-factor binding/unbinding and temporal changes in DNA supercoiling, among others \cite{sanchez2013genetic, jones2018bursting}.

\subsection{Continuous, non-Markovian models of gene expression}
 
Another improvement in experimental resolution, which necessitated a theoretical revision, was the ability to measure mRNA levels at an accuracy finer than a whole molecule, i.e., quantify the amounts of different parts of the same polymeric mRNA \cite{chen2015genome, wang2019measuring}. While the models discussed above depict mRNA creation and elimination as point processes, this is, in fact, a rather poor approximation, especially in bacteria, where the timescale of synthesizing a full-length mRNA is comparable to the lifetime of these molecules, both typically on the order of several minutes \cite{chen2015genome}. Consequently, much of the mRNA in the bacterial cell is expected to consist of partial rather than full-length molecules. mRNA number is thus better approximated as a continuous, rather than discrete, variable. In addition, its stochastic kinetics cannot be assumed to be Markovian, but instead exhibit finite memory of transcription initiation events. Writing the master equation for mRNA dynamics becomes more challenging, but still possible, using various heuristic approaches \cite{xu2016stochastic, jiang2021neural}. Furthermore, once mRNA kinetics becomes tractable at sub-molecular resolution, other intricacies of the gene expression process, which were ignorable earlier, reveal themselves. These include the complex coupling between multiple co-transcribing RNAPs, between them and the ribosomes translating the same transcript, and between those ribosomes and the enzymes degrading the mRNA molecule \cite{kim2019long, chen2015genome, kim2018ribosome}. While the theoretical consideration of mRNA kinetics at sub-molecule resolution is an important direction for future work, it is outside the focus of this Colloquium.

\section{The kinetic effects of gene replication}
\label{chapter-generep}

\subsection{Beyond the static picture}

``The dream of a bacterium is to become two bacteria”, said Francois Jacob \cite{jacob}. In other words, cells beget more cells, and, at least in the realm of unicellular organisms, they typically do so as rapidly as they can. One consequence of this is that the attractor of any cellular observable in an exponentially growing population is not stationary but rather cyclo-stationary, i.e., a limit cycle (called, appropriately, the {\it cell cycle})  in which a new cell is born, and during a finite period doubles its volume and the number of all cellular components. These components are then partitioned into two (approximately) identical daughter cells, et cetera.  

The constant-rates models introduced in Chapter \ref{chapter-static} ignore all these features. These models are stationary — the number of gene copies is held constant at one, reaction rates are unchanging, and the attractors, too, are stationary. Evidently, these models must be revised to reliably capture gene expression in growing, proliferating cells. We now embark on the construction of these revised models, doing so in a gradual manner. In this chapter, we consider the impact of a single, discrete event: the replication of the encoding gene. As we will see, this doubling (referred to as a change in ``gene dosage") creates a time varying gene expression pattern along the cell cycle. In the next chapter, we will shift our focus to the continuous aspects of cell growth and examine the modifications that this growth imposes on gene expression.

In discussing the kinetic effects of gene replication, We focus on mRNA, rather than protein, levels. The reason is that as the step immediately downstream of the DNA, transcription responds first, and more dramatically, to the discontinuous change in dosage. The effect on protein levels is delayed and, owing to the longer lifetime of proteins, temporally smoothed (recall Fig. \ref{fig:bulk-msrmnt}). In addition to these differences in kinetics, our experimental ability to follow transcription along the cell cycle currently outpaces that for protein kinetics (we return to this point in Section \ref{subsection_experimental_observations}). While earlier single-cell measurements were still ignorant of the cell cycle phase of individual cells, and hence had to contend with mapping dosage changes to expression ``noise" (see Box \ref{box-noise}), more recently it has become possible to measure how mRNA numbers vary with cell-cycle progression \cite{wang2019measuring, pountain2022transcription}. As we will see, these new studies reveal diverse patterns, some involving non-monotonous changes along the cell cycle. We discuss the possible interpretations of these empirical findings.

\subsection{Replication of the gene of interest}

\begin{figure}[htp]
    \centering
    \includegraphics[width=8cm]{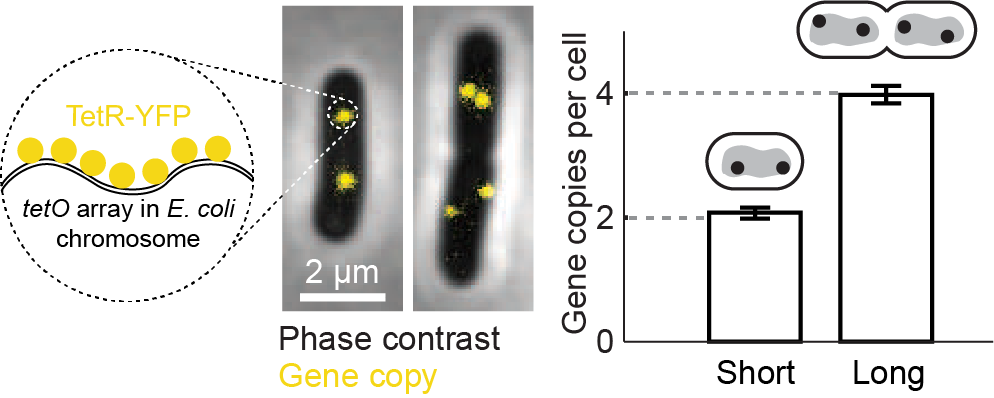}
    \caption{Gene dosage varies twofold within a population of growing bacteria. Left, the genomic region of a specific gene in the {\it E. coli} chromosome is fluorescently labeled in live cells. Right, the number of gene copies in cells about to divide (``long") is double that of newborn cells ("short"). Figure adapted from \cite{sepulveda2016}. Reprinted with permission from AAAS.}
    \label{fig:ls-genecopies}
\end{figure}

Single-cell measurements of gene expression became prevalent at the beginning of the new millennium \cite{elowitz2002stochastic,ozbudak2002regulation}, and drove extensive utilization of stochastic models for the process. These models typically followed the constant-rates formulation of Chapter \ref{chapter-static}. This was because the single-cell measurements at the time were ignorant of the age of the individual cells (i.e., its cell cycle phase), thus ``legitimizing" the exclusion of this critical observable from the models. Nevertheless, because the data was typically acquired from asynchronous populations of growing cells, gene dosage was expected to vary twofold within the population (Fig. \ref{fig:ls-genecopies}). 
To incorporate this feature into the models of gene expression, several studies \cite{so2011general,jones2014promoter} used the assumption that mRNA levels will rapidly equilibrate to reflect the new gene dosage resulting from replication. In that case, a population of growing cells is considered to be composed of two subpopulations, of cells before and after gene replication. The size of each subpopulation will depend on the growth rate and the genomic position of the gene\footnote{In the case of rapidly growing bacteria, where multiple replication rounds take place simultaneously, even the newborn cells may have more than one gene copy, Fig. \ref{fig:ls-genecopies}.}. Once these parameters are known, they can be used to calculate the contribution of dosage heterogeneity to both the mean and the noise in gene expression (see Box \ref{box-noise}). This procedure has the advantage of allowing the use of a standard constant-rates gene expression model (Chapter \ref{chapter-static}) for the interpretation single-cell experiments, where the cell cycle phase of individual cells is unknown.   

\noindent\fbox{%
    \parbox{\textwidth}{%
        \textbf{Box \ref{box-noise} Extrinsic versus Intrinsic noise}
\label{box-noise}     
        
As we discussed in Chapter \ref{chapter-static}, beyond the ensemble-averaged dynamics one is often interested in predicting the cell-to-cell variance in expression. Cell growth and replication will contribute to this variance. For example, the rate of mRNA production in Eq. (\ref{ODE1}), $k_m$, will vary between cells depending on their age (cell cycle phase) due to the difference in gene copy-number before and and after replication, as well as changes in the copy number of RNAP and other molecular players involved in gene expression. Beyond the changes in production rate, asymmetric partitioning at cell division will also contribute to cell-to-cell variability. One way of representing all these different contributions to heterogeneity is by treating them as sources of  ``extrinsic noise", in addition to the ``intrinsic noise" associated with the stochasticity of the kinetic scheme itself model \cite{swain2002intrinsic, huh2011non, peterson2015effects, jones2014promoter}, as we shall now explain.

Consider the copy-number of a given mRNA or protein in the cell, within the class of stochastic models introduced in Chapter \ref{chapter-static}, albeit when the transcription and translation rates $k_m$, $k_p$ are time-dependent, reflecting their potential change along the cell cycle. The law of total variance enables one to decompose the variance of a random variable $Y$ (e.g., $M$ or $P$) into two components, by conditioning on another variable or set of variables $X$ \cite{blitzstein2015introduction,hilfinger2011separating, fu2016estimating}: 
\be Var[Y] = E[Var[Y|X]] + Var[E[Y|X]].\ee

The first term on the RHS corresponds to taking the variance of the variable (say, protein level) when \textit{conditioning} on the parameters (in our case, $k_m$ and $k_p$), and then taking the expectation value over the probability distribution of these parameters. This term corresponds to the noise we expect in the simpler models with fixed rates. This is known as \textit{intrinsic} noise. In the case of the two-state model, for example, it can be shown to scale as $(\bar{b}+1) \bar{P}$, where $\bar{b}$ is the burst size of Eq. (\ref{burst_size}) and $\bar{P}$ is the mean expression level of the protein in question (as expected, the standard deviation, normalized by the mean expression level, will be smaller for highly expressed proteins). The second term in the decomposition accounts for the fluctuations in the reaction rates, and is known as \textit{extrinsic noise}; For any given set of parameters we are conditioning on, consider the mean (the expectation value) of the variable of interest (e.g., protein level). The extrinsic noise is simply the variance of that expectation value over the distribution of the varying parameters. 

The expected contribution to the extrinsic noise in gene expression from several cell-cycle features--e.g., gene doubling, cell growth and division--has been calculated \cite{swain2002intrinsic, jones2014promoter, peterson2015effects, huh2011non}. 
The challenge, however, is that these noise contributions may eventually dominate over the intrinsic component reflecting the kinetic scheme---which is often our main interest. This limitation of noise-based analysis thus motivates directly measuring the cell-cycle phase of individual cells, and explicitly considering factors such as cell volume and gene dosage in the analysis of gene expression.       

    }%
}

 One assumption underlying the treatment above is that transcription rate is proportional to gene dosage, hence would double once the gene replicates. However, experimental data indicates that this simple proportionality is sometimes violated. Specifically, mRNA production rate may be a sublinear function of gene copy number, a feature referred to as ``dosage compensation". 
 Dosage compensation may be advantageous to the organism as a means of buffering expression against the unavoidable change of gene copy-number during cell growth. 
 The subject is outside the premise of this work (reviewed in \cite{bar2016dealing}).

 Notwithstanding the possibility of dosage compensation, the approximation that mRNA levels instantaneously track gene dosage relies on the assumption that mRNA lifetime (which determines the adaptation time to dosage doubling) is negligible compared to the cell generation time. This, however, is again a poor assumption in rapidly growing bacteria, where the two timescales may be within a few-fold of each other. In that case, the temporal kinetics of mRNA (as described, e.g., by Eq. (\ref{ODE1})) must be solved, while matching the two boundary conditions: before and after replication of the gene (at time ${\it t}_r$), and before and after cell division (at time ${\it t}_d$). One arrives at the following expression for the population averaged mRNA number over time, $\overline{M}(t)$ \cite{peterson2015effects}:
 


\be
\overline{M}(t) = 
  \begin{cases}
     \frac{k_m}{\gamma_m}(1-\frac{e^{-\gamma_m(t-t_r+t_d)}}{2-e^{-\gamma_mt_d}}) & 0 < t \leq t_r \\
     \frac{2k_m}{\gamma_m}(1-\frac{e^{-\gamma_m(t-t_r)}}{2-e^{-\gamma_mt_d}}) & t_r < t < t_d
  \end{cases}
\label{peterson-ODE}\ee 

As expected, the finite mRNA lifetime results in a smoothed (filtered) response to the discontinuous change in gene dosage (see Fig. \ref{fig:mengyu-pr} below). Beyond the population mean, the contribution to extrinsic noise in mRNA numbers can also be calculated for this case, using the approach described in Box \ref{box-noise} \cite{peterson2015effects} .

\subsection{Cell-cycle dependent transcription: Experimental observations} 
\label{subsection_experimental_observations}
\begin{figure}[htp]
    \centering
    \includegraphics[width=8cm]{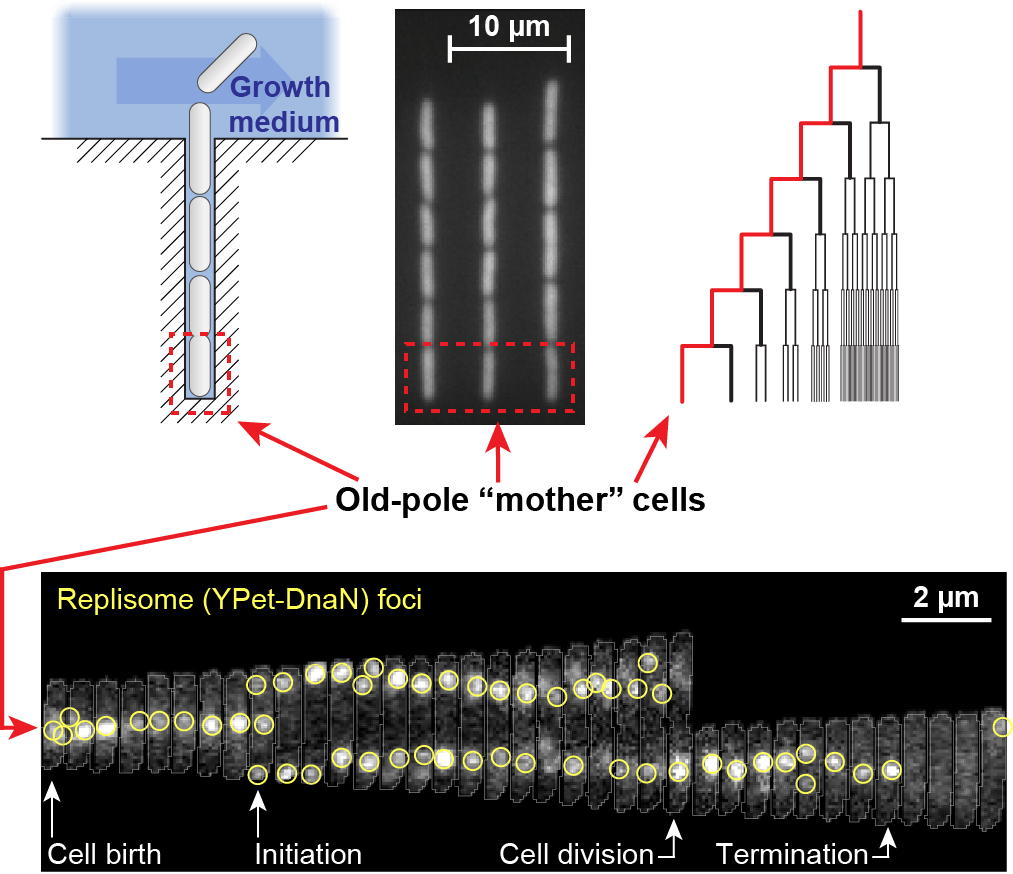}
    \caption{Tracking the progression of the bacterial cell cycle in real time. Top, the ``Mother Machine" microfluidic device enables high-throughput observation of mother cells (adapted from \cite{wang2010robust}, with permission from Elsevier). Bottom, the progression of genome replication in {\it E. coli} is followed by fluorescently labeling the cellular replication machinery, or ``replisome" (adapted from \cite{knoppel2023regulatory}, reproduced under Creative Commons Attribution License 4.0 (CC BY)).    
     }
    \label{fig:mother-machine-data}
\end{figure}

Bacterial cells can nowadays be grown and tracked under the microscope for many generations \cite{wang2010robust}. By combining bright field and fluorescence microscopy, key events during the cell cycle--e.g., the initiation or termination of genome replication--can be detected in each cell and their timing recorded (Fig. \ref{fig:mother-machine-data}). But while this approach has allowed an empirical characterization of the bacterial cell cycle, the real-time measurement of mRNA and proteins production along the cell cycle remains largely outside the current experimental capability. Multiple challenges of live-cell microscopy contribute to this problem. For one, long-term measurement of gene expression relies on the detection of fluorescent proteins. To emit their signal, these proteins must undergo a slow and stochastic process of fluorophore maturation, which lowers the temporal resolution of the measurements to multiple minutes at best \cite{balleza2018systematic}. Consequently, the inference of cell-cycle expression patterns has remained a challenge (but see \cite{rosenfeld2005gene, hensel2015cell, zopf2013cell} for exceptions). In some instances, the dependence on fluorescent protein maturation can be circumvented by devising a detection scheme that relies on the change in localization of preexisting proteins rather than production of new ones. This approach was used successfully to detect and count both mRNA (Fig. \ref{fig:bursting} above) and gene loci (Fig. \ref{fig:ls-genecopies} above) in live bacteria, and analogous schemes have been proposed for detecting translation \cite{wu2016translation}. However, issues of detection sensitivity, resolution, and perturbation to cell physiology must be resolved before robust, long-term measurement of gene expression along the bacterial cell cycle becomes possible.       

\begin{figure}[htp]
    \centering
    \includegraphics[width=5cm]{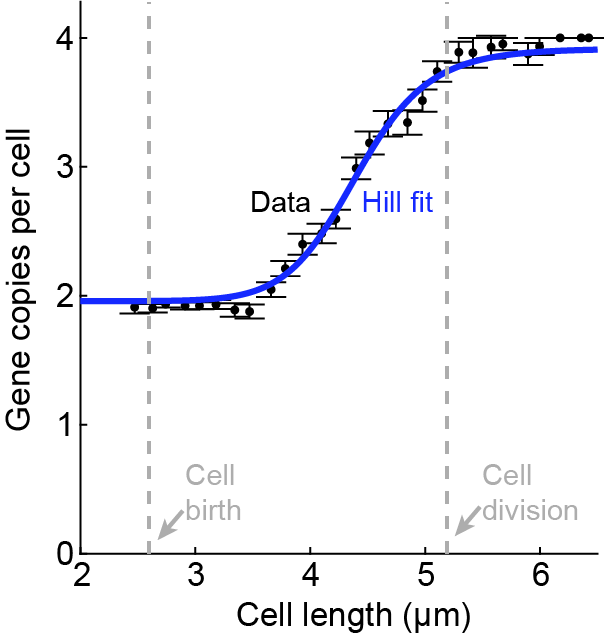}
    \caption{Cell length approximates cell-cycle progression in {\it E. coli}.  Bacteria were chemically fixed and imaged to measure cell length and the copy number of a given genomic locus, following \cite{wang2019measuring}. The gene dosage doubles sharply, as expected. Data by Tianyou Yao (unpublished).}
    \label{fig:mengyu-fros}
\end{figure}

As an alternative to tracking gene activity in live cells, the details of cell-cycle dependent transcription can be revealed by analyzing snapshots of chemically fixed cells. This approach leverages the fact that for exponentially growing cells, cell size (in rod-shaped bacteria like {\it E. coli}, its length) can be approximately mapped to its age (cell cycle phase). This is demonstrated by the distribution of measured cell lengths, which reflects the expected statistics of cell ages during exponential growth \cite{pountain2022transcription}, as well as the measured gene copy-number, which exhibits a step-like change as a function of cell length, see Fig. \ref{fig:mengyu-fros} \cite{wang2019measuring}. An additional advantage offered by size-based analysis is that, in {\it E. coli}, the initiation of genome replication is triggered, on average, at a given cell volume rather than age \cite{wallden2016synchronization,ho2018modeling, zheng2016interrogating}, thus making size a natural axis along which to examine the effect of this event.

\begin{figure}[htp]
    \centering
    \includegraphics[width=6cm]{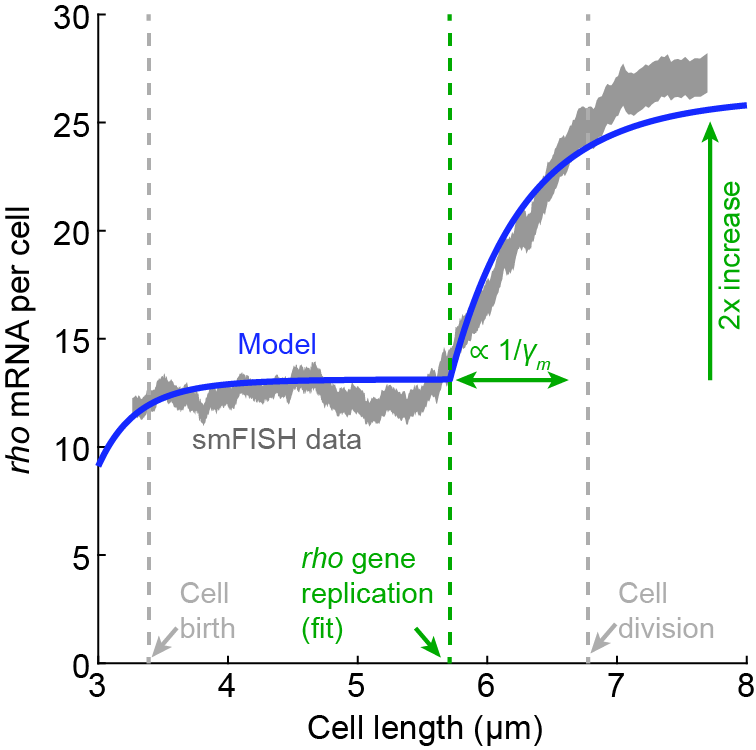}
    \caption{Transcription of a strong {\it E. coli} promoter matches the prediction of a simple mRNA-follows-dosage model. mRNA copy number for the gene {\it rho} was measured using smFISH. The single-cell data was binned by cell length (shaded curve). Blue line, fit to Equation \ref{peterson-ODE}, with time mapped to cell length as described in \cite{pountain2022transcription}. }
    \label{fig:mengyu-pr} 
\end{figure}

\begin{figure}[htp]
    \centering
    \includegraphics[width=8cm]{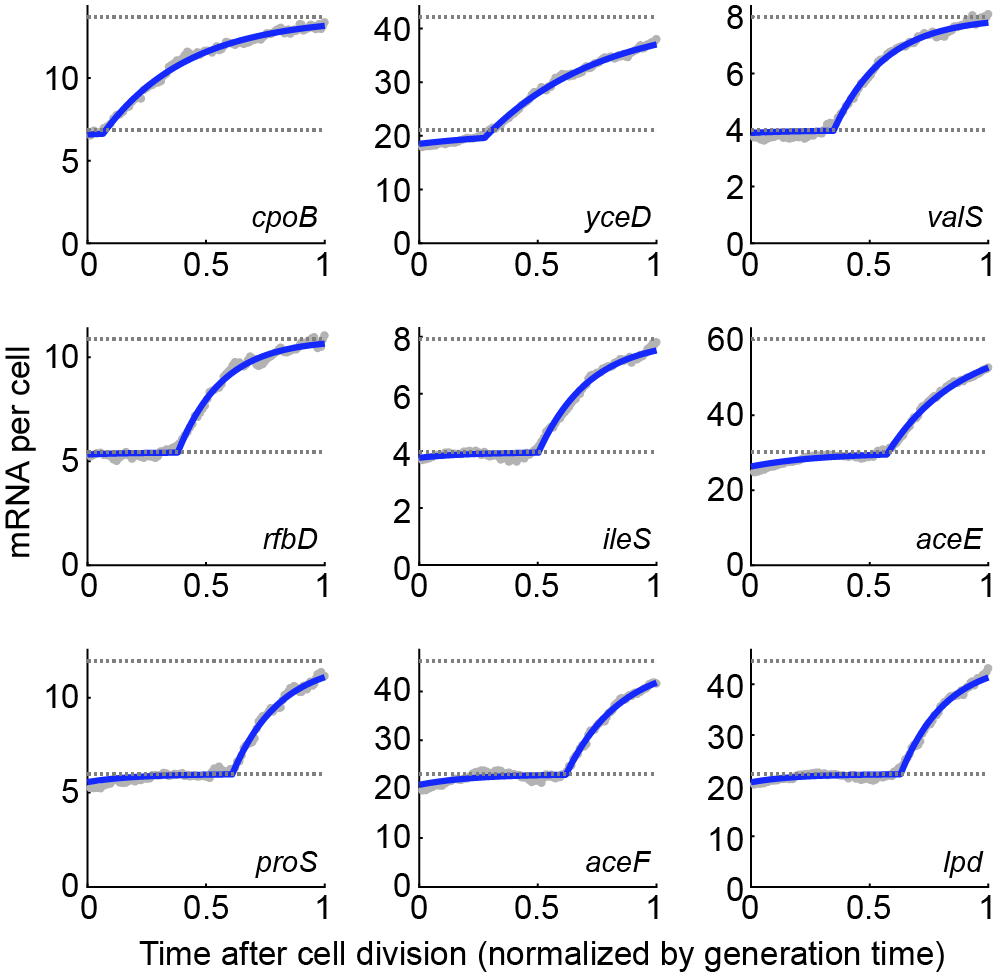}
    \caption{Single-cell RNA sequencing (scRNA-seq) analysis reveals an mRNA-follows-dosage pattern across multiple {\it E. coli} genes. scRNA-seq expression, converted to mRNA copy-number, is plotted against cell age. Horizontal dotted lines indicate the expected steady-states levels before and after gene replication. Blue line, fit to Equation \ref{peterson-ODE}. Data from \cite{pountain2022transcription}. Additional analysis by Kevin McDonald and Tianyou Yao.}
    \label{fig:andrew-canonical}
\end{figure}

Using this approach to measure the mRNA level along the cell cycle for several strongly expressed promoters in {\it E. coli} (Fig. \ref{fig:mengyu-pr}) revealed good agreement with the model of Eq. (\ref{peterson-ODE}), where mRNA levels track gene dosage with a finite adaptation period. While the imaging-based method is limited to quantifying only a few promoters at a time, a recent study introduced an algorithm for sorting the full transcriptome of individual cells (obtained using single-cell RNA sequencing, scRNA-seq) along the cell cycle, thus opening the door to identifying the cell-cycle expression pattern across the whole genome \cite{pountain2022transcription}. The sequencing-based results agree well with imaging approach, and, for many {\it E. coli} genes, reveal a similar mRNA-follows-dosage pattern along the cell cycle, again captured well by Eq. (\ref{peterson-ODE}) (Fig. \ref{fig:andrew-canonical}).

\subsection{Non-monotonic transcription patterns: observations and possible mechanisms}

\begin{figure}[htp]
    \centering
    \includegraphics[width=6cm]{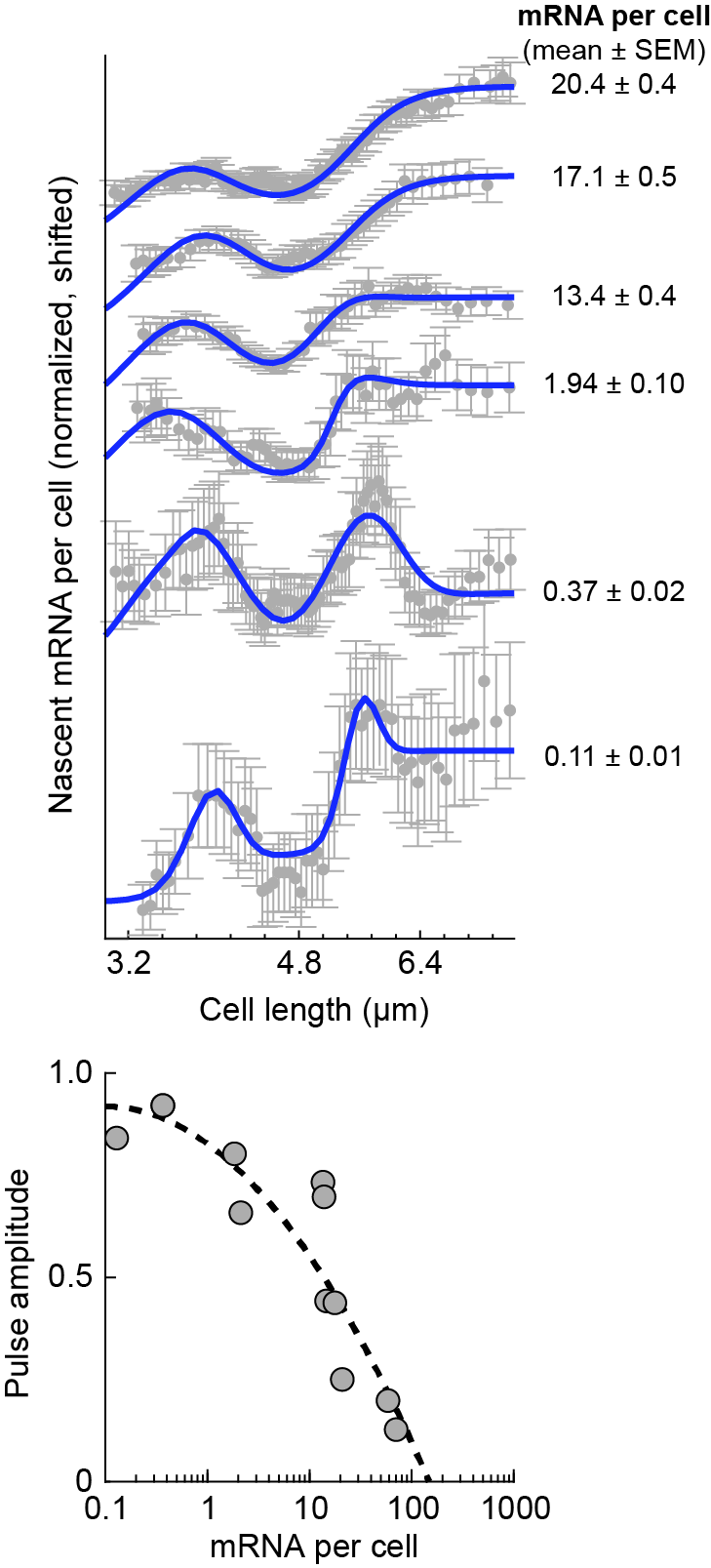}
    \caption{As mRNA expression decreases, cell cycle dependence shifts to a non-monotonic pattern, with a transient peak around the time of gene replication. Top, smFISH measurement of nascent mRNA from the ${\textit{lac}}$ promoter at different expression levels. Bottom, the effect of gene replication on ${\textit{lac}}$ transcription as a function of the expression level. The amplitude of transient expression pulse (relative to the expected dosage response) is plotted versus the mRNA copy number per cell. Adapted from \cite{wang2019measuring}.}   
    \label{fig:mengyu-plac}
\end{figure}

\begin{figure}[htp]
    \centering
    \includegraphics[width=8cm]{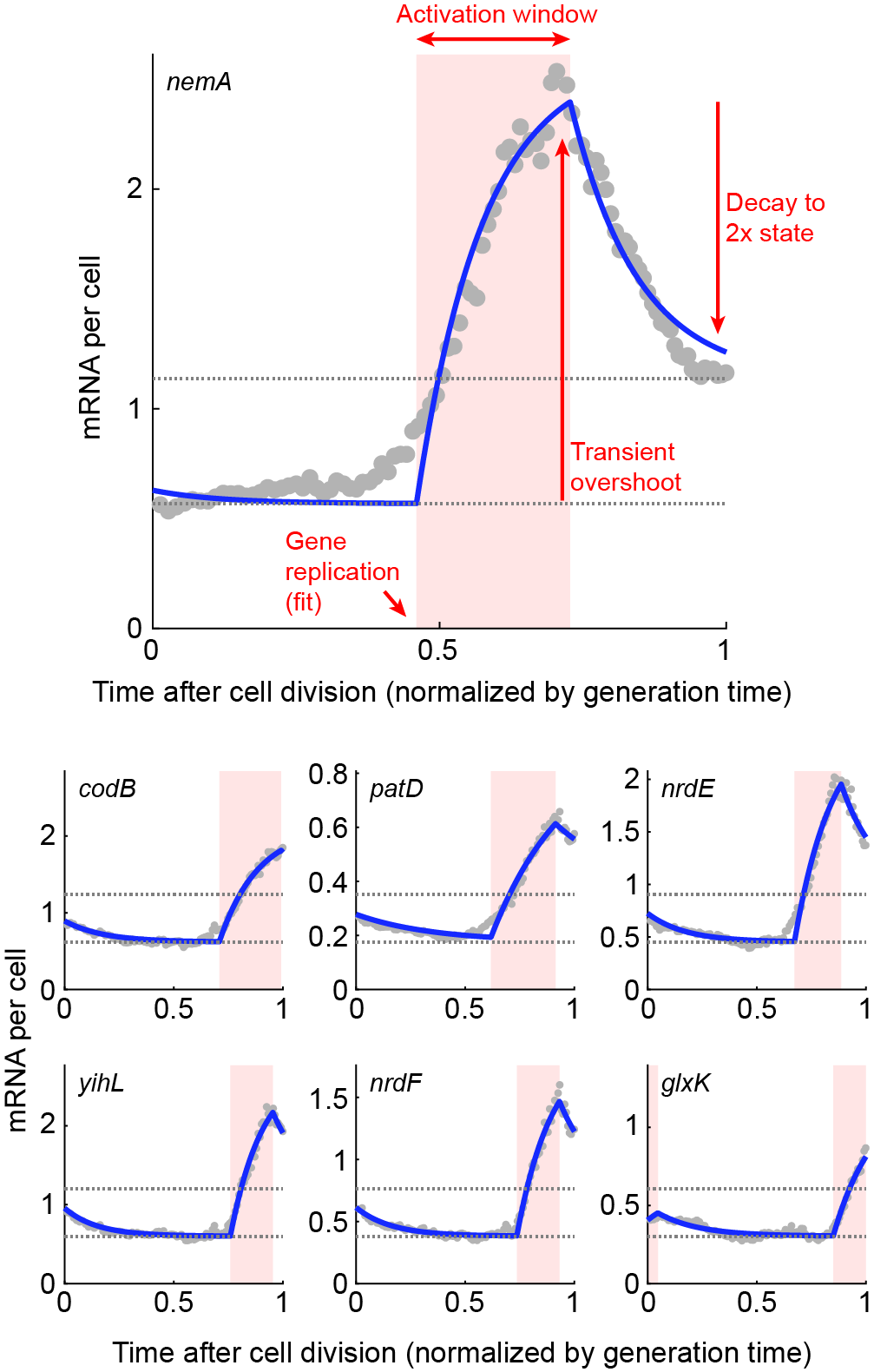}
    \caption{Single-cell RNA sequencing (scRNA-seq) analysis reveals non-monotonic cell-cycle dependence for multiple {\it E. coli} genes. scRNA-seq expression, converted to absolute mRNA copy-number, is plotted against the normalized cell age. Horizontal dotted lines indicate the expected steady-states levels before and after gene replication, demonstrating the failure of the simple mRNA-follows-dosage model. Blue line, fit to a model where gene replication is accompanied by a transient increase in promoter activity. Data from \cite{pountain2022transcription}. Additional analysis by Kevin McDonald and Tianyou Yao.}
    \label{fig:andrew-noncanonical}
\end{figure}

While multiple promoters examined exhibit the step-like pattern expected from the simple dosage-dominated picture, many other promoters show non-monotonic changes in mRNA level along the cell cycle, where the expected increase accompanying gene replication is both preceded and followed by a {\it decrease} in expression (Fig. \ref{fig:mengyu-plac}) \cite{wang2019measuring}. The anecdotal observations using microscopy are again reflected in the RNA sequencing analysis, which suggests that many {\it E. coli} genes exhibit this behavior (Fig. \ref{fig:andrew-noncanonical}). These non-monotonic expression patterns, whose origin is currently unknown, provide an opportunity for testing some of the current ideas regarding the drivers of gene expression in growing cells. Here we briefly discuss two classes of hypotheses.

{\it 1. Replication-triggered transcription.} An idea long discussed in the bacterial literature is that transcription of low-expression proteins takes place preferentially around the time of gene replication rather than with a uniform probability along the cell cycle \cite{guptasarma1995does, golding2019revisiting}. The idea is premised on the many conceivable ways in which passing of the DNA replication machinery through the gene may transiently increase transcription, beyond the obvious change in dosage discussed above. The hypothesized effects include changes to DNA topology (supercoiling) ahead and behind the replicated gene; changes in the spatial position of the gene in the cell during replication, making it more accessible to RNAP and ribosomes; and the displacement of bound transcription factors acting as repressors by the replication process---their removal resulting in transient transcriptional activity until they rebind \cite{guptasarma1995does, golding2019revisiting}. Consistent with the latter hypothesis, the activity patterns of the repressor-controlled ${\textit{lac}}$ promoter 
appears to gradually shift, from the canonical dosage-tracking at high expression to a pulsatile replication-associated one, as the repression by the transcription factor LacI tightens (Fig. \ref{fig:mengyu-plac}). 
Beyond the various biophysical effects of DNA replication, it is conceivable that the doubling of dosage itself may elicit a non-monotonic transcriptional response. Gene replication creates a step-like change in dosage, and the resulting effect on transcription can be seen as the ``step response" of the genetic circuit of which the replicated gene is part. Depending on the topology of that circuit, e.g., the presence of one or more negative feedback loops \cite{milo2002motifs}, this response may be non-monotonic or even oscillatory \cite{stricker2008fast}.  


{\it 2. RNAP competition during genome replication.} 
In Chapter \ref{chapter-contgrowth} we discuss the idea that transcription of a given gene is limited by RNAP availability, and thus that the different genes compete for this limited resource. The competition could conceivably result in non-monotonic transcription along the cell cycle, reminiscent of the empirical data of Fig. \ref{fig:mengyu-plac} and \ref{fig:andrew-noncanonical}: rising when the gene-copy doubles, but diminishing at other times while the rest of the genome replicates, since this replication produces competing targets for RNAP. The expected activity pattern is complicated by the elaborate scheme of genome replication in rapidly growing bacteria, where multiple, nested replication events run simultaneously and the rate of new-genome (hence RNAP targets) production varies along the cell cycle \cite{frederick1990physiology, wallden2016synchronization}.  



Identifying the mechanistic origins of cell-cycle dependent transcription requires characterizing mRNA numbers across the genome and as a function of multiple parameters,  including the cell's growth rate and each promoter's expression level and regulatory topology, interrogation that is now becoming possible thanks to emerging single-cell transcriptomics approaches based on imaging \cite{dar2021spatial} and sequencing \cite{pountain2022transcription}. 

\section{A self-consistent model for gene expression in growing cells}
\label{chapter-contgrowth}




In considering cell growth and reproduction, we focused above on a single element, the gene of interest, and examined the consequences of its replication for the expression of the corresponding mRNA. But this discrete replication event takes place as part of the doubling of cell volume and all cellular components, including the entire genome and the gene expression machinery. What is the effect of these global changes? To answer this question, we consider a simplified picture in which the cell undergoes continuous growth, while putting aside for now genome replication and cell division. As we will see, this level of description enables us to identify important constraints on the rates of transcription and translation in growing cells. 


\subsection {Correcting for cell growth by normalization by cell volume} \label{norm-volume}

 
As noted above, early studies of stochastic gene expression utilized a constant-rates formulation (Chapter \ref{chapter-static}). The change in gene dosage during cell growth, rather than modeled explicitly, was mapped to an added noise term (BOX \ref{box-noise}). But, of course, not only the gene of interest but {\it all} genes, as well as mRNAs and proteins, are expected to vary twofold within a population of growing cells (even ignoring the additional stochastic effects). A common way of addressing this heterogeneity became to normalize protein numbers (mRNA measurements were still uncommon at the time) by the cell volume, i.e., consider \textit{concentrations} rather than molecular numbers, and interpret these numbers in a constant-rates framework \cite{elowitz2002stochastic,thomas2021coordination,jia2021concentration}.
Intuitively, this normalization should---at least to some extent---take care of the doubling of all cellular components during the cell cycle: for cells undergoing binary fission (such as {\it E. coli}) cellular concentrations of genes, mRNAs, and proteins are expected to be identical before and after cell division\footnote{This could break down for proteins that are partitioned not according to the ratio of volumes \cite{lin2019optimal,min2021transport}.}. To render this argument more rigorous, let us use Eqs. (\ref{ODE1})-(\ref{ODE2}) to derive the temporal dynamics of the cellular concentrations of mRNA and protein, $m$ and $p$. This requires us to explicitly consider cell growth, since volume increase inherently leads to the dilution of both species. Denoting by $\P$ the protein copy-number, we have $p=\P/V$. Therefore, the dynamics of the ensemble-averaged concentration follows\footnote{For simplicity, we omit the $\overline{x}$ notation used earlier for ensemble means.}
\be \frac{d\p}{dt} = \frac{1}{V}\frac{d\P}{dt} - \frac{\P}{V^2}\frac{dV}{dt}. \label{dilution1} \ee
Let us assume that, at a given point along the cell cycle, proteins are produced at a rate $\kappa$, which is potentially time-dependent (within the models of Chapter \ref{chapter-static}, $\kappa$ will be proportional to the instantaneous mRNA copy number). Assuming exponential growth of cell volume, with rate $\lambda = \frac{1}{V}\frac{dV}{dt}$, we find:
\be \frac{d\p}{dt} = \frac{\kappa}{V} - (\lambda+\gamma_p) \p. \label{dilution2}  \ee 
We thus need to account for the effect of dilution by adding $\lambda$ to the protein degradation rate $\gamma_p$. The possible time dependence of $\kappa$ allows for the possibility of a time-independent protein concentration as a stationary solution. In particular, if $\kappa(t) \propto V(t)$, we see that the additional term on the RHS will allow for a fixed point of the dynamics, even in the absence of protein degradation: a homeostatic concentration level, at which dilution balances production. As we will discuss in Section \ref{experiments_for_phases}, this is consistent with experimental observations. 

\subsection{Constraints on the rates of transcription and translation}
The preceding discussion left undetermined $\kappa(t)$, the time-dependent rate of protein production in Eq. (\ref{dilution2}). To determine it, we will need to develop a more complete model, which considers both transcription and translation \textit{simultaneously} in the context of continuous cell growth. This section introduces such a model.

\subsubsection{The rate of protein production}
\label{growth_law_hwa}
We first digress from the preceding discussion of stochastic gene expression, and briefly recap a celebrated ``growth-law" observed experimentally in growing microbes, including \textit{E. coli} and budding yeast: as nutrient conditions are varied, one finds that the fraction of total protein mass in the cell taken by ribosomes increases linearly with the cellular growth rate \cite{scott2010interdependence, metzl2017principles}. There has been considerable work recently related to this growth law, outside the scope of the current Colloquium (see \cite{scott2023shaping}). Here, we will ignore many of the subtleties and highlight the simple qualitative rationale for the observed behavior, which will become pertinent to our discussion of gene expression. 

Since ribosomes produce all proteins within the cell -- including other ribosomes -- a coarse-grained model for their auto-catalytic production (neglecting degradation) would suggest:
\be \frac{dR}{dt}  = k_{tr} \tilde{\Phi}_r R , \label{growth_law}\ee
where $R(t)$ is the total number of ribosomes in the cell\footnote{We typically think of the rates of biochemical reactions as determined by the cellular \textit{concentrations} of molecules. Why have we switched back to working with \textit{copy numbers}? The scenario to have in mind is one where ribosomes are the limiting cellular resource for protein production, and once a ribosome completes translation of a given mRNA, it is only idle for a brief moment, after which it randomly encounters another mRNA and begins translating again. We can thus assume that all ribosomes are producing proteins at any given time, arriving at Eq. (\ref{growth_law}). Of course, the equation can be rewritten in terms of concentrations, but the derivation and interpretation are simpler when working with copy numbers. This will also be true for the subsequent derivations in this chapter.
(In reality, only a finite fraction of ribosomes are active at any moment \cite{metzl2017principles}. However, this will only introduce a prefactor of order unity into the equations, reflecting this fraction.)},
$k_{tr}$ is the translation rate (per ribosome), and $\tilde{\Phi}_r$ is fraction of ribosomes that are actively translating ribosomal proteins. Note that although each ribosome is composed of tens of smaller proteins, in this coarse-grained, simplified equation the ribosome is treated as a single, self-replicating entity.  Furthermore, we have neglected the fact that each ribosome has a large RNA component (in addition to proteins). However, one may argue that producing ribosomal RNA is much ``cheaper" than producing ribosomal proteins, as evidenced by the fact that in \textit{E. coli} the fraction of RNAP in the proteome (a few percents, typically) is considerably lower than fraction of ribosomes. A more systematic discussion of these features is presented in \cite{reuveni2017ribosomes}.
The solution of Eq. (\ref{growth_law}) is exponential growth of the ribosomal copy number with a rate proportional to $\tilde{\Phi}_r$, which we also expect to equal the growth rate of the cell. Note that since ribosomes produce all other proteins as well, those proteins' numbers will also increase exponentially, and the fraction of ribosomes in the cell will equal $\tilde{\Phi}_r$.  
We conclude that the growth rate should be proportional to the fraction of ribosomes in the proteome, thus providing a possible explanation for the experimentally observed ``growth-law". 

At the heart of the simple model above lies the assumption that ribosomes are limiting for translation, with the protein production rate proportional to their copy number. This contrasts with the models of Chapter \ref{chapter-static}, where changes in the rates of protein production were only associated with changes in mRNA copy numbers, while the ribosomal levels played no explicit role. Furthermore, since ribosomal numbers are expected to increase as the cell grows, the protein production rate would not be constant in time, in contrast to the assumptions of the earlier models. What should the constant rates in, e.g., Eqs. (\ref{ODE1})-(\ref{ODE2}), be replaced with to be consistent with the ribosomal growth law?

To answer this question, recall that the ribosome-centric model we introduced corresponds to a scenario where ribosomes are always ``hungry", and are actively translating some mRNA at any moment in time \cite{lin2018homeostasis}. Within this picture, the mRNAs corresponding to different genes compete for ribosomes' attention, and the protein production rate for gene $i$ will depend not on the absolute level of the corresponding mRNA, but rather on its relative abundance in the pool of mRNAs, which will in turn determine the chance that the next ribosome to become available will encounter it. Under this scenario, the protein production rate reads:
\be \frac{dP_i}{dt} = k_{tr} R \frac{\tilde{M}_i}{\sum_j \tilde{M}_j}, \label{translate}\ee
with $\tilde{M}_j$ the \textit{effective} mRNA copy number of gene $j$ -- also accounting for its affinity for ribosome binding -- and the summation is over all genes in the genome. For simplicity, below we neglect the heterogeneity in ribosomal binding affinity, and hence associate $\tilde{M}_j$ with the actual mRNA copy number, $M_j$. The total protein production rate (i.e. the production rate summed over all proteins) will be, by construction, limited only by the ribosome number and independent of the mRNA levels. Those mRNA levels, however, dictate the relative rates of protein production for different genes. To determine these rates, we thus turn next to the laws governing \textit{transcription} within the ribosome-centric model of cell growth. 


\subsubsection{The rate of mRNA production}
\label{section-rnap-comp}

In analogy to the preceding discussion regarding translation, let us assume that the transcription rate of each gene is limited by the (time dependent) cellular number of RNA polymerases (RNAPs), which we denote by $N$. Similarly to what we previously assumed of ribosomes, we envision that RNAPs (or, as before, a finite fraction of them) are always busy transcribing, with the rate of mRNA production from a given gene dictated by the fraction of RNAPs actively transcribing that gene. Which specific gene is transcribed by the next available RNAP will depend on the particular gene's copy number and the affinity of RNAP to the promoter region of the gene. The propensity to transcribe can be further modulated  by the action of \textit{transcription factors}, proteins that bind the genome and affect gene expression by, e.g., sterically blocking the site RNAP should bind to \cite{ptashne2002genes}. To incorporate these combined effects, we write an expression analogous to Eq. (\ref{translate}), where the fraction of mRNAs corresponding to a particular gene determined its translation rate. Here, the corresponding quantity is one we term the {\it gene allocation fraction}:
\be \Phi_i = \frac{g_i}{\sum_j g_j}, \label{gene_alloc}\ee
where $g_i$ is a coarse-grained quantity that reflects the copy number of a given gene as well as the regulatory features above -- in other words, $g_i$ determines how competitive a given gene is in capturing RNAP's attention. The mRNA copy number then obeys:
\be \frac{dM_i}{dt} = k_{tx} N \Phi_i - \gamma_m M_i, \label{transcribe} \ee
with $k_{tx}$ the transcription rate (per RNAP)\footnote{Not to be confused with $k_m$ of Eq. (\ref{model1}).}, $N$ the number of RNAPs, and $\gamma_m$ the mRNA degradation rate (which, for simplicity, we will assume to be identical for all genes). 

Before proceeding to solve equations (\ref{translate}) and (\ref{transcribe}), it is important to note that, in both of these equations, one of the indices {\it i} corresponds to the genes encoding ribosomes, and another to those encoding RNAP. Also note that the translation rates in Eq. (\ref{translate}) depend explicitly on the mRNA levels, which in turn are given by Eq. (\ref{transcribe}) -- provided that we know the time-dependent RNAP level $N(t)$. 
 
\subsection{Solving the model}
\label{solving_model_continuous}


Eqs. (\ref{translate}) and (\ref{transcribe}) comprise a closed set of equations for the production of both mRNA and proteins -- including ribosomes and RNAPs themselves. These equations were written under the explicit assumptions that ribosomes (rather than mRNA copy numbers) limit the overall rate of protein production in the cell, and, similarly, RNAPs (rather than the gene copy number) limit the overall rate of RNA synthesis. We will later relax these assumptions, and, in doing so, Eqs. (\ref{translate}) and (\ref{transcribe}) will come to describe one regime (later denoted as ``Phase I") out of several possibilities described by the continuous growth model. 

To proceed, consider Eq. (\ref{translate}) for the case of genes encoding ribosomal proteins. It reads \cite{lin2018homeostasis}:
\be \frac{dR}{dt} = k_{tr} R \frac{M_R}{\sum_j M_j}, \label{rib}\ee
with $M_R$ the copy number of ribosomal mRNA. Next, we note that the solution of Eq. (\ref{transcribe}) for the mRNA of gene {\it i} is given by:
\be M_i(t) = e^{-\gamma_m t} M_i(0)+  e^{-\gamma_m t} k_{tx} \frac{g_i}{\sum_j g_j} \int_0^t e^{\gamma_m t'} N(t') dt'. \label{Mrna_dynamics} \ee
At long times compared with the mRNA lifetime, the initial condition for the mRNA copy number will not matter, and we conclude that:
\be M_i/M_j = g_i / g_j.  \label{mrna_allocation}\ee
Plugging this into Eq. (\ref{rib}) leads to a closed equation for the ribosome number:
\be \frac{dR}{dt} = k_{tr} \Phi_r R , \label{rib2}\ee
reproducing the functional form of the ``growth law" of Eq. (\ref{growth_law}). Previously, $\tilde{\Phi}_r$ was defined as the fraction of active ribosomes translating ribosomal proteins. Here, $\Phi_r$ is the gene allocation fraction -- ``hardcoded" into the DNA since it depends on the gene copy number and promoter strength (but also, potentially, on the modulation by transcription factors). To see why these two quantities are identical, note that, according to Eq. (\ref{mrna_allocation}) the gene allocation fraction results in an identical mRNA fraction, which, in turn, implies (according to Eq. (\ref{rib})) the same ribosomal fraction.

Eq. (\ref{rib2}) for the ribosome number is closed, hence $R(t)$ is now known, and predicted to be exponential. This allows us to revisit Eq. (\ref{translate}), but consider the expression of other proteins within the cell, finding along a similar vein:
\be \frac{dP_i}{dt} = k_{tr} \Phi_i R , \label{protein_exp}\ee
the solution of which gives us:
\be P_i(t) = P_i(0)+ \frac{k_{tr}}{\lambda} R(0) \Phi_i [e^{\lambda t}-1], \label{protein_dynamics} \ee
with $\lambda =  k_{tr} \Phi_r $ the growth rate obtained from Eq. (\ref{rib2}).

It is useful to recast these equations in terms of \textit{concentrations}, which will help reveal that the behavior we obtain indeed represents a steady-state (i.e., homeostasis) in terms of those cellular concentrations. This follows the logic of our discussion around Eq. (\ref{dilution2}), but now the protein production rate $\kappa(t)$ is obtained explicitly. Let us assume that the \textit{total} protein concentration in the cell is fixed to a value $c$, independent of the changes to cell volume during growth. In Chapter \ref{chapter-concentration} we will discuss how such homeostasis may be achieved, but for now we can take it as an empirical observation \cite{crissman1973rapid,kubitschek1983buoyant, rollin2023physical}. From Eq. (\ref{protein_exp}) we find that:
\be \frac{dp_i}{dt} = k_{tr} \Phi_i r - \lambda p_i   , \label{OU} \ee
with $p_i$ denoting protein {concentration} and $r$ the ribosome concentration.

\begin{figure}[htp]
    \centering
    \includegraphics[width=8cm]{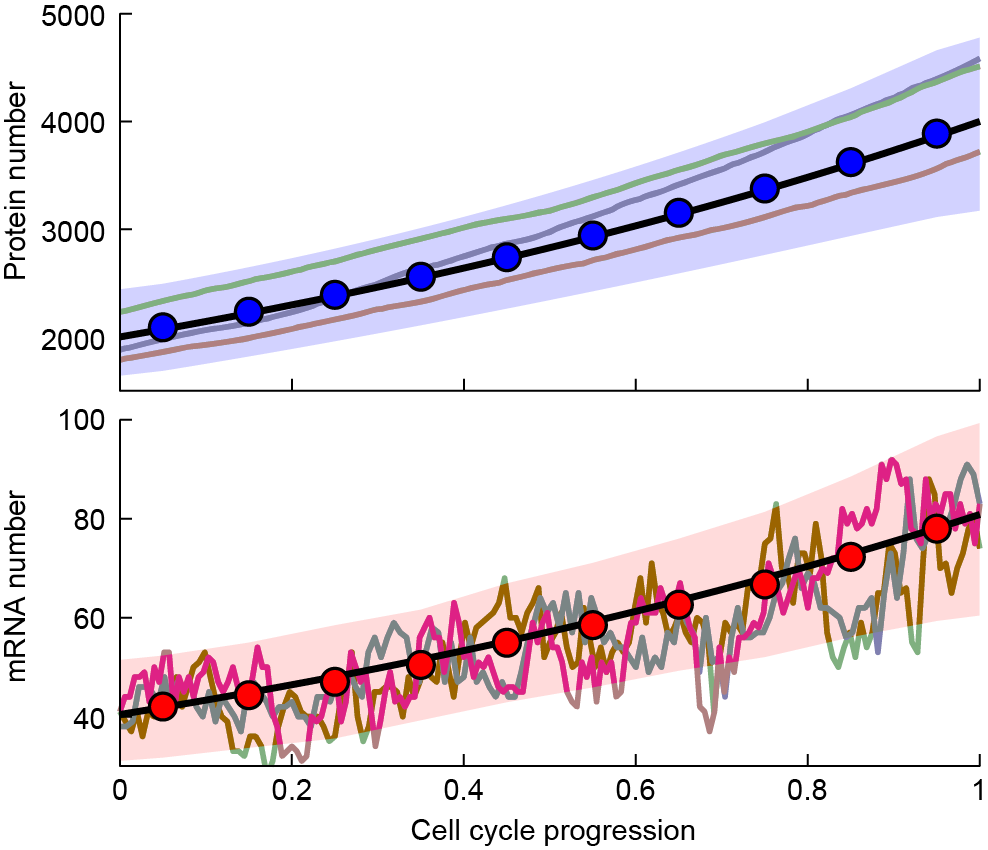}
    \caption{ Fluctuations of mRNA and proteins levels within the continuous growth model. 
    The results of Gillespie (stochastic) simulations of the ribosome-centric gene expression model, corresponding to Eqs. (\ref{translate}) and (\ref{transcribe}), are shown. The protein and mRNA levels increase exponentially with time, with strong fluctuations in the mRNA and much weaker ones for the proteins. The background shows three individual trajectories of the stochastic dynamics, while the circles show the mean of 130 cell cycles (with the colored bands representing the standard deviation).  The black lines are the theoretical predictions of exponential growth. 
    Adapted from \cite{lin2018homeostasis}, and reproduced with permission. }
    \label{fig:gillespie}
\end{figure}

According to this equation, the concentration of each protein behaves as the position of an overdamped particle in a harmonic potential -- it is subject to a linear restoring force, attracting it to the steady-state concentration of $c \Phi_i$. Note that since we have written an ODE for the ensemble-average, stochasticity has been neglected; adding it would lead to small fluctuations around the steady-state solution, as shown in Fig. \ref{fig:gillespie}.

\begin{figure}[htp]
    \centering
    \includegraphics[width=8cm]{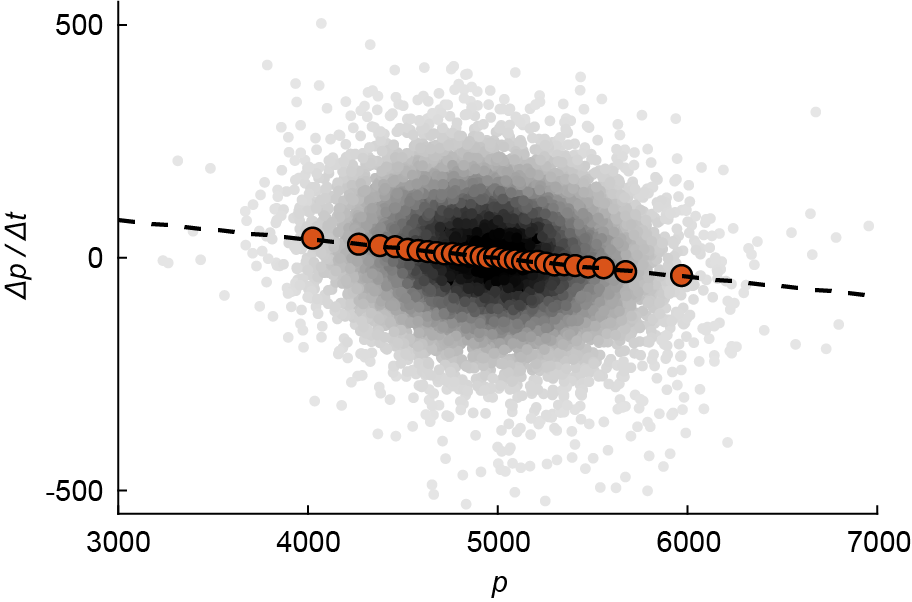}
    \caption{An effective restoring force on protein concentration.
    The continuous growth model predicts that the time-derivative of protein concentration is linearly dependent on the deviation of the concentration from its steady-state value, see Eq. (\ref{OU}). The figure shows experimental data in \textit{E. coli} (from Ref. \cite{tanouchi2017long}), where the concentration of a constitutively expressed (i.e., unregulated) gene, fluorescently labeled, is tracked over time. The red points are binned single-cell data: the time-derivative of the concentration is averaged over many cells with approximately the same concentration, thus suppressing the fluctuations and revealing the underlying linear trend. Adapted from Ref. \cite{lin2021disentangling}, with permission.}
    \label{fig:OU}
\end{figure}

With stochasticity present, the mechanical analogy with a particle in a harmonic potential essentially maps the dynamics of the protein concentration to the well-known Ornstein-Uhlenbeck process,  corresponding to a confined particle subject to Brownian noise, and described by a simple Langevin equation \cite{amir2020thinking}. Without the restoring force, the particle would diffuse to infinity. Without the noise, the overdamped particle would reside in a particular ``coordinate" (corresponding to the homeostatic concentration $c \Phi_i)$. With both features present, there exists an equilibrium solution -- in our problem, a stationary distribution for the concentration, centered around its fixed point in the absence of noise. In fact, from observing the fluctuations involved in gene expression, we may infer the relative contributions of intrinsic and extrinsic noise (discussed in box \ref{box-noise}) acting on a particular gene: In the absence of any extrinsic noise (and assuming no additional regulation), Eq. (\ref{OU}) predicts that the strength of the ``restoring force" is the growth rate $\lambda$. Extrinsic noise can be shown to weaken this restoring force \cite{lin2021disentangling}.  The experimental data for \textit{E. coli}, corroborating the prediction of an effective, linear restoring force, is shown in Fig. \ref{fig:OU}. Note that the fluctuations due to the stochastic term, which will supplement Eq. (\ref{OU}), are what enables us to measure this restoring force -- since without these fluctuations the cellular concentration would have been perfectly constant. This is reminiscent of the way in which natural variability between cells enables one to draw conclusions regarding cell size control \cite{amir2018learning, ho2018modeling, kar2023using}.

An analogous equation can be written for the mRNA concentrations:
\be \frac{dm_i}{dt} = k_{tx} c \Phi_n \Phi_i -(\lambda +\gamma_m) m_i, \label{mrna_fixed}\ee
with $\lambda$ the growth rate.
Considering again \textit{numbers} instead of concentrations, Eqs. (\ref{OU}) and (\ref{mrna_fixed}) tell us that at long times compared with the cell's doubling time, protein and mRNA copy numbers will increase exponentially (since concentrations are stationary and volume increases exponentially)\footnote{Note that since multiple cell divisions occur during this time, we should consider the total protein numbers in all the progeny of the initial cell considered.}. This conclusion -- namely, exponential growth, with identical rates, of protein and mRNA numbers -- is robust to the introduction of stochasticity and cell divisions into the continuous growth model, as is illustrated in Fig. \ref{fig:gillespie}.  


The main simplifying assumption in the derivation above is the constancy of the effective fraction of gene copy number for each gene, $\Phi_i$. This implied that the relative fraction of RNAPs transcribing any two genes does not change in time. Consequently, the relative mRNA levels corresponding to these genes (at times long compared with the mRNA lifetime) are also given by $\Phi_i$, as are the resulting relative protein levels. In reality, of course, the DNA encoding all genes is replicated during cell growth. How would this affect the above calculation? If the duration of replicating the entire genome is small compared to the cell's doubling time, as is sometime the case for eukaryotes, the relevant fractions $\Phi_i$ would remain the same before and after the replication of the entire genome, and, assuming also that mRNA lifetime is short, the predictions above would hold. In fast growing bacteria, on the other hand, we are in a very different regime -- replicating the genome typically takes a considerable fraction of, or even  
\textit{longer} than, the birth-to-division time, and the resulting existence of multiple replications simultaneously further complicates the picture above \cite{frederick1990physiology}. This scenario has not yet been studied in depth within the class of models presented above. It may, conceivably, give rise to the non-monotonic gene expression patters discussed in chapter \ref{chapter-generep}, since after a particular gene is replicated (leading to a fast rise in its mRNA levels), the subsequent replication of additional genes, together with the effects of competition for the limiting RNAPs (as reflected in Eq. (\ref{transcribe})) is expected to result in a \textit{decrease} in the mRNA levels of the gene in question.

\subsection{Revisiting the assumptions -- what limits transcription and translation?} 
\label{sect-revisit}

In constructing the model for gene expression in growing cells, we made specific assumptions about the factors limiting gene expression: RNAP for transcription, ribosomes for translation\footnote{A priori, one may imagine that protein production and cell growth will also be limited by the rate of transporting nutrients into the cell, which would depend on the surface area to volume ratio, hence on cell size and cell-cycle progression. Empirically, however, this does not appear to be the case: Studies in both mammalian cells \cite{mu2020mass} and \textit{E. coli} \cite{zheng2016interrogating} found that even a dramatic perturbation to cell size did not alter cells' growth rate.}.
 We now explore the consequences of relieving these assumptions. As noted, doing so leads to different dynamics for mRNA, protein, and cell growth, defining other regimes (or "phases") of the continuous growth model. Later, in Section \ref{experiments_for_phases}, we refer to various experimental studies and attempt to assess in which phase of the continuous growth model cells of various organisms reside.  

First, we modify the premise of the original model by assuming that transcription is limited by the DNA amount in the cell, rather than RNAP availability as we posited initially (protein production is still assumed to be ribosome-limited). As motivation for studying this case, consider a gedanken experiment---we will later discuss an actual experiment of this sort---where cell volume continually increases as before, but the amount of DNA remains fixed. The limiting resource for transcription is initially assumed, as before, to be RNAP, resulting, in accordance with Eq. (\ref{mrna_fixed}), in a constant cellular concentration of proteins. Cellular DNA, on the other hand, is gradually diluted. It is evident that, at some stage, DNA rather than RNAP will become limiting for transcription: a single DNA template would be insufficient to support transcription in an enormous cell. 

To understand how this would come about mechanistically, we may consider the limit where the volume/DNA ratio is sufficiently large such that RNAPs are packed to their limit on the DNA; clearly, there is a physical limit to the number of RNAPs that can fit on any particular region of the DNA, in turn limiting transcription.  
Reaching the physical limit of RNAP occupancy is, however, not the only possibility. Ref. \cite{lin2018homeostasis} arrives at similar results by considering, instead, stochastic RNAP kinetics: binding/unbinding at the promoter, and the initiation of  transcription when bound. The authors show that when the free RNAP concentration is low, the model reduces to that of Section \ref{solving_model_continuous} (i.e., RNAPs are limiting), but that this inevitably breaks down for large volume/DNA ratio, at which the amount of DNA becomes limiting for transcription.

\begin{figure}[htp]
    \centering
    \includegraphics[width=8cm]{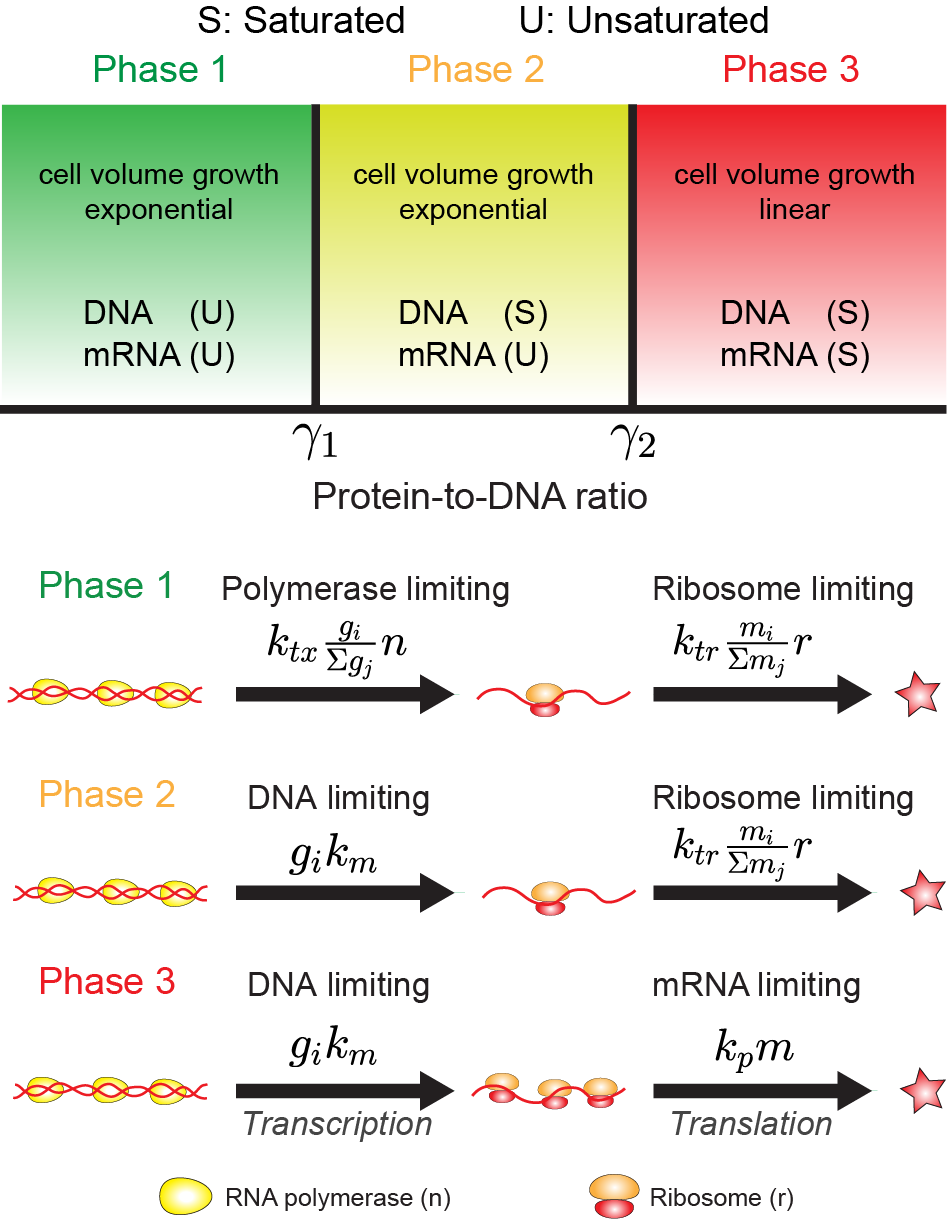}
    \caption{The different phases of the continuous growth model. The figure shows, schematically, the three phases of the model, where protein and mRNA production rates are limited by different resources: transcription is limited by RNAP number \textit{or} by the DNA template, and translation is limited by ribosome number \textit{or} by the mRNA numbers. The reaction rates listed in the bottom panel refer to the protein and mRNA \textit{concentrations} rather than copy numbers. 
     Adapted from \cite{lin2018homeostasis}, and reproduced with permission.}

    \label{fig:phase_diagram}
\end{figure}

Regardless of the underlying mechanism, when DNA becomes limiting for transcription, mRNA production follows:
\be \frac{dM_i}{dt} = g_i k_m - \gamma_m M_i . \label{transcribe2} \ee
The total amount of cellular mRNA depends on the DNA level in the cell, and will saturate to a constant rather than increase exponentially. However, importantly, so long as gene dosage stays unchanged the \textit{relative} amounts of mRNAs between different genes will still obey:
\be \frac{Mi}{M_j} = \frac{g_i}{g_j}= \frac{\Phi_i}{\Phi_j}, \ee
with $g_i$ the effective gene copy numbers and $\Phi_i$ the gene-allocation fraction, defined in Eq. (\ref{gene_alloc}). Since protein production is still described by Eq. (\ref{translate}), and depends on the relative amounts of mRNAs, the previous predictions of the ribosome-centric model for the \textit{protein} production remain intact. In particular, protein levels still increase exponentially in time, and the ribosome growth law of Eq. (\ref{growth_law}) remains valid. We refer to this regime, where transcription is DNA limited and translation is ribosome limited, as ``Phase II" of the continuous growth model, see Fig. \ref{fig:phase_diagram}.

What happens if we further relax the assumption that ribosomes are limiting for protein production? In that case, the model becomes identical to the constant-rates model (aside from the effects of gene dosage due to DNA replication, as discussed in Chapter \ref{chapter-generep}), and protein accumulation becomes \textit{linear} rather than exponential in time (in the absence of protein degradation, which would lead to its saturation at a finite value). We refer to this as ``Phase III" of the continuous growth model. The three regimes are summarized schematically in the ``phase diagram" of Fig. \ref{fig:phase_diagram}. Phase I is the regime analyzed in Section \ref{solving_model_continuous}, where RNAP is limiting for transcription and ribosomes for translation. Phase II is the regime where ribosomes are still limiting translation (rather than the mRNAs) but transcription is limited by the DNA template rather than RNAP. In Phase III, as in the constant-rates model of Chapter \ref{chapter-static}, DNA is limiting transcription while mRNAs are limiting translation\footnote{For realistic parameter values, as one increases the volume/DNA ratio, the transition to the regime where transcription is limited by DNA rather than RNAP occurs \textit{before} the regime where mRNAs become limiting for translation -- thus preventing a 4th phase where ribosomes are not limiting but RNAPs are \cite{lin2018homeostasis}.}.

\subsection{The limiting resource for transcription and translation: Experimental evidence}
\label{experiments_for_phases}
The analysis above indicated that the identity of the limiting factors for transcription and translation will result in different temporal dynamics of mRNA and protein levels. What does the experimental data suggest for different organisms? We begin by reviewing results for protein levels, then proceed to discuss mRNA.

\subsubsection{The scaling of protein levels with time and cell volume}
The question of how protein levels scale with time or cell volume is a long-standing one. Already in the 1970's, work based on radioactive labeling showed that, in certain mammalian cells, protein numbers are proportional to cell volume \cite{crissman1973rapid}. More recently, by flowing cells through a microfluidic device embedded in a cantilever, and measuring the latter's resonance frequency, the buoyant mass of growing cells (which is typically dominated by proteins \cite{frederick1990physiology,hosios2016amino}) was measured to the remarkable precision of a picogram -- 1 \% of the mass of a typical \textit{E. coli} cell (for mammalian cells, the relative accuracy is an order of magnitude higher).
The signal is precise enough that the time-derivative of the mass can be evaluated. For linear growth in time of the biomass, this derivative is expected to be constant, while for exponential growth it will be proportional to the instantaneous cell mass. Data from four different system -- the bacteria \textit{E. coli} and \textit{B. subtilis}, the budding yeast {S. cerevisiae}, and mammalian cancer cells -- was inconsistent with linear growth but consistent with exponential growth \cite{godin2010using, cermak2016high}. In fission yeast, recent experiments found that the rate of protein production was approximately proportional to cell volume,  as would be expected if mass and volume grow exponentially \cite{basier2023cell}.
The observation of approximately exponential growth hints that cells are in either Phase I or Phase II of the continuous growth model, in which ribosomes are limiting for translation. As noted in Section \ref{growth_law_hwa} above, in the case of bacteria, the aforementioned ``growth law" relating ribosome concentration to growth rate is, too, suggestive that ribosomes are limiting for protein production, thus reinforcing this conclusion.

Note that, in phases I and II of the continuous growth model, if the duration of DNA replication is short compared to the cell cycle, total protein production will be exponential (since the protein allocation of all proteins -- including ribosomes -- will be identical before and after DNA replication). While the gene dosage effects discussed in chapter \ref{chapter-generep} impact the expression levels of a given gene (both in terms of transcription and translation), in phases I and II the total ribosome copy number is the only determinant of protein production. As long as an approximately constant fraction of ribosomes is devoted to ribosome production, their fraction in the proteome will remain constant and total protein production will be exponential. This is a reasonable approximation for many eukaryotes, since, as we mentioned, the duration of DNA replication may constitute a small fraction of the cell cycle, which within the continuum growth model implies constant protein allocations. But why this would be valid for bacteria such as \textit{E. coli}, where the duration of DNA replication is comparable to the cell cycle duration, is not obvious. How can we explain then the approximately exponential growth of biomass measured in this case? Could there be deviations from exponential growth of mass that cannot be revealed using the current experimental setups? (analogous to those observed for mammalian cells \cite{mu2020mass}). Alternatively, the tight control exerted over ribosomal levels \cite{frederick1990physiology} might enable bacterial cells to maintain a constant ribosome fraction in the proteome throughout the cell cycle, leading to true exponential growth of biomass. 


\subsubsection{The scaling of mRNA levels with time and cell volume}
In a similar manner, we may consider the change in mRNA levels with cell cycle progression, a question we began examining in Chapter \ref{chapter-generep}. In the current context, it is convenient to consider the scaling with cell volume. In phase I of the continuous growth model, mRNA level is proportional to cell volume (since both are exponential in time). Importantly, the linear dependence between the two is agnostic as to the level of DNA, thus the same scaling will exist before and after DNA replication. This picture contrasts with that obtained in phases II/III of the model, where DNA replication is reflected in a twofold jump in mRNA levels, preceded and followed by a plateau -- absent in Phase I, but consistent with the patterns we saw in Chapter \ref{chapter-generep} for many {\it E. coli} genes. 

In contrast to the bacterial behavior, experimental data for mammalian cells \cite{padovan2015single} shows a clear linear dependence between mRNA copy number of a given gene and cell volume, consistent with the expected Phase I behavior under the assumption that RNAP limits transcription. In fission yeast, the experimental evidence supports the same conclusion: by studying mutants with differing cell size, it was found that global mRNA levels correlated with the RNAP occupancy (the fraction of RNAP bound to the promoter region) in a manner consistent with the above picture, where the polymerases form the limiting factor for transcription
\cite{zhurinsky2010coordinated}. Recent work revealed that transcription rates in fission yeast scaled approximately linearly with cell volume, also consistent with this interpretation \cite{basier2023cell}. Alternative evidence, also supporting the RNAP-limiting picture in the same organism, was recently provided by experiments using single-molecule mRNA counting \cite{sun2020size}, which found a linear relation between mRNA number and cell size. Similar behavior was reported in budding yeast \cite{SWAFFER2023}. However, this latter work suggested that the mRNA lifetime -- which we so far assumed to be constant -- also changes throughout the cell cycle, compensating for the sublinear dependence of the RNAP occupancy with cell size, and together leading to the linear relation.

The linear scaling between mRNA copy number and volume, reported for the evolutionarily distant mammalian cells, fission, and budding yeast, hints at the possibility of universal behavior. Nonetheless, as we saw in Chapter \ref{chapter-generep}, similar results were not reported for bacteria, in which diverse behavior is observed, and where matters are 
potentially complicated by the fact that the duration of DNA replication is comparable to the birth-to-division time, such that the gene-allocation fraction changes throughout the cell cycle. 

A recent study measured both transcription and translation levels as a function of growth-rates in \textit{E. coli}, albeit in bulk measurements rather than the single-cell level \cite{balakrishnan2022principles}. While that study was thus unable to identify in which phase of the continuous growth model bacteria reside, the results nevertheless confirm several of the general predictions of the model: across the genome, there was a strong, linear correlation between the mRNA and protein levels of different genes, as is expected from Eqs. (\ref{mrna_allocation}) and (\ref{protein_dynamics}). This behavior is expected in all phases of the continuous growth model, but is not obvious a priori, and is violated, for example, in mammalian tissues where the assumption of continuous growth does not hold \cite{harnik2021spatial}. 

\subsubsection{Experiments observing the slowdown of exponential growth}
\begin{figure}[htp]
    \centering
    \includegraphics[width=8cm]{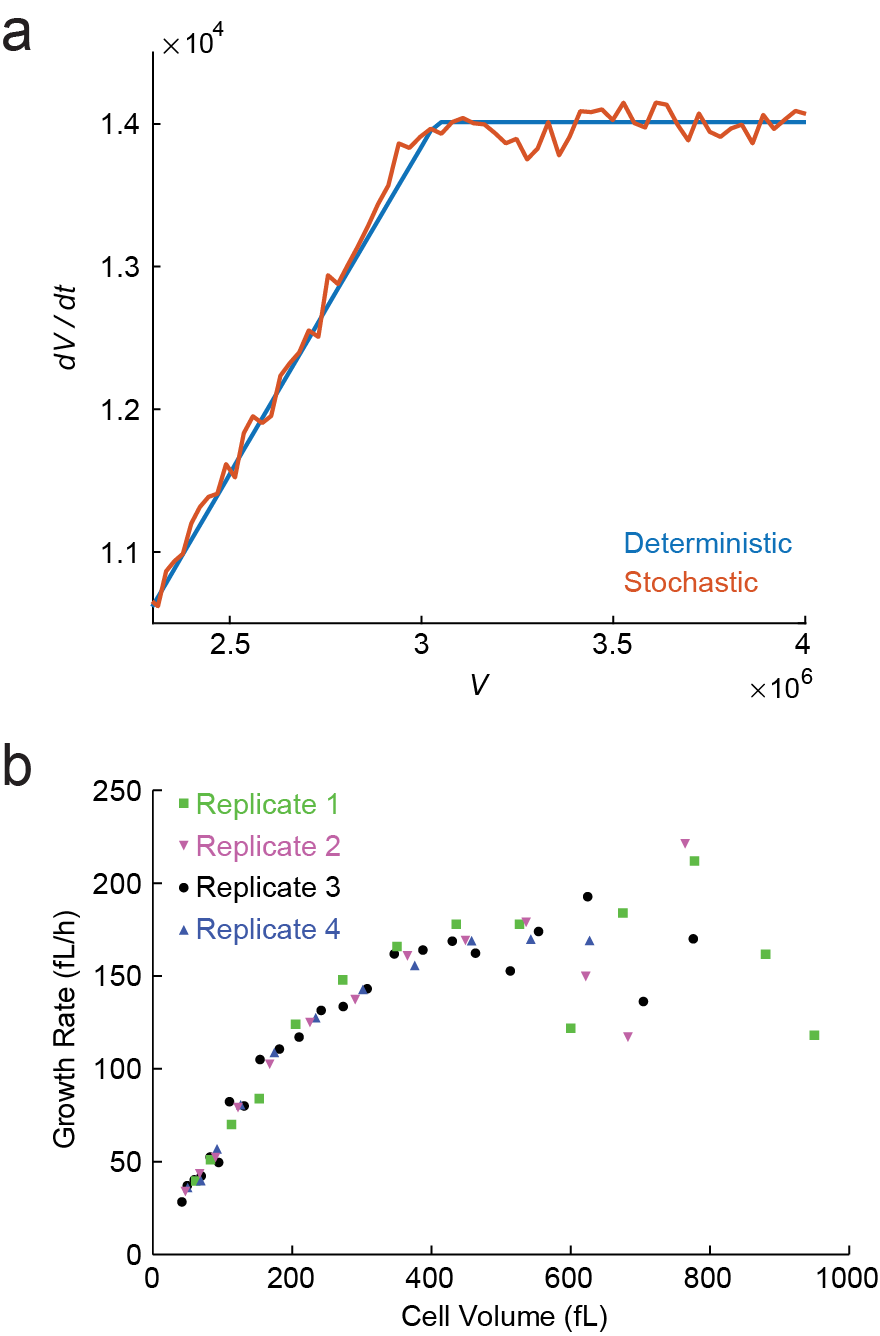}
    \caption{The continuous growth model predicts a transition from exponential to linear growth of cell mass. (a) As the volume/DNA contents increases (e.g., in a growing cell where DNA replication and cell division are blocked), the continuous growth model predicts a transition from Phase I/II to Phase III, implying a change from exponential to linear growth. The plot, adapted from Ref. \cite{lin2018homeostasis}, shows both the deterministic calculations as well as the results of Gillespie simulations of the stochastic model. (b) Experiments on budding yeast, where cells are arrested in the pre-replication (G1) phase, and hence keep growing without division or DNA replication. This leads to abnormally large volume/DNA ratios, and cells manifest a transition from exponential to linear growth of cell volume, as predicted in (a). Adapted with permission from Ref. \cite{neurohr2019excessive}.}
    \label{fig:amon}
\end{figure}

As we saw above, in Phase II of the continuous growth model cell growth is exponential in time, whereas in Phase III it is linear. In fact, the continuous growth model predicts a sharp transition from exponential to linear growth of cell volume as the volume/DNA ratio increases, see Fig. \ref{fig:amon}. This is consistent with results in budding yeast \cite{neurohr2019excessive}, where abnormally large cells can be formed by blocking DNA replication and cell division while cell growth continues essentially unperturbed. A sharp transition is found between an exponential and a linear growth regime (Fig. \ref{fig:amon}), and it appears that the transition occurs at a critical value of volume/DNA -- consistently with the results discussed above. Moreover, it was found that cells with a larger copy number of the DNA (ploidy) manifested the transition at a larger volume, approximately proportional to the number of chromosomes -- as expected from the model. Other experiments suggested that in mammalian cells, too, growth rates decline when cells grow too large without increasing their DNA content \cite{zatulovskiy2022delineation,liu2022beyond}.


\section{Outlook}
\label{chapter-concentration}

We began this Colloquium by discussing some of the simplest gene expression models. We illustrated how analyzing the fluctuations in gene expression provides important insights and helps reveal shortcomings of models that, by capturing the ensemble-averaged observables, initially appeared to perform well. To incorporate the impact of cellular growth, we then surveyed models that account for the replication of the genome and its effects on gene expression, as well as models that consider the constraints imposed by the finite amounts of the key cellular machinery -- in particular, RNAPs and ribosomes. The models were compared with experimental data from various organisms, and were able to explain some of the data -- with some outstanding puzzles remaining.

In this chapter, we first briefly describe recent extensions of the continuous growth model coupling gene expression with cell growth. One important remaining limitation of these models is that they cannot explain the constancy of the global protein density (i.e., the density of the total protein mass within the cell), which therefore still needs to be invoked as an explicit assumption. We discuss recent experimental and theoretical works addressing this gap. Finally, all of the models discussed in this Colloquium assumed that the cytoplasm is well-mixed and spatial effects play no role. We conclude by highlighting the evidence for the role of spatial effects in gene expression.

\subsection{Extensions and generalizations of the continuous growth model}
\label{section-extensions}

As mentioned in section \ref{sect-revisit}, one mechanism which can interpolate Phase I and Phase II of the continuous growth model is the kinetics of RNAP binding and unbinding from the gene promoter. If one assumes identical kinetic rates for all of genes, the protein levels (for all genes) are found to scale linearly with cell volume -- consistent with the calculations of Section \ref{solving_model_continuous}. Recent work relaxed this assumption, and considered the heterogeneity in the binding strength of different promoters \cite{wang2021heterogeneous}. Interestingly, it was found that within this extended model the expression of some genes scales \textit{sub-linearly} with the volume, while others scale \textit{super-linearly}, consistent with experimental findings in budding yeast \cite{chen2020differential}. Note that this would imply that the proteome composition would change with cell size: the super-linear genes would be over-represented in larger cells, and vice-versa. This is indeed observed experimentally in human cell lines \cite{lanz2022cell,cheng2021size}.

Another recent work extended the ``phase diagram" discussed in the previous chapter, finding a new regime where the limiting factor for translation is neither ribosomes nor mRNA transcripts alone, but rather the formation of ribosome-mRNA complexes \cite{calabrese2023total}. One motivation for developing this model were experiments where yeast cells transcribed mRNAs that were quickly degraded before being translated. As the amount of such mRNA was increased, the cellular growth rate declined \cite{kafri2016cost}. This result is surprising in light of the model presented earlier, where it was assumed that transcription is ``cheap" and the burden on growth rate rises from protein production rather than transcription. If, instead, one considers a regime where ribosome-mRNA complexes are limiting, this behavior arises naturally: there would be a growth ``cost" associated with producing mRNAs, even when they are not successfully translated.

Finally, in Chapter \ref{chapter-contgrowth}, the ``gene allocation fraction" played a crucial role in determining the transcription rate (and through it, the protein production rate), but we neglected the possible effects of transcription factors (TFs). In fact, it is not difficult to include those in the model, as was done in Ref. \cite{guo2021exploring}. Conceptually, since the model considers explicitly all proteins in the cell, one needs to account for the TFs, and the term $k_{tx}$ in Eq. (\ref{transcribe}) depends (non-linearly) on the concentration of these TFs. Ref. \cite{guo2021exploring} shows that adding such regulatory elements would, at some critical level of TFs, lead to a destabilization of the gene regulatory network, destroying the concentration homeostasis we derived in Chapter \ref{chapter-contgrowth} (and leading to chaotic behavior instead). 

\subsection{Concentration homeostasis in growing cells}
\label{section-concentration}
In previous chapters, we discussed models where transcription and translation occur at a constant rate (independent of volume), we well as and ones where different genes and mRNAs ``compete" with each other for the attention of the relevant molecular machinery. The latter class of models led naturally to \textit{relative} concentrations which are maintained over time, fluctuating around a well-defined value. However, an explicit assumption we made was that the \textit{global} concentration within the cell is maintained: we assumed, based on previous experimental evidence in both bacteria and mammalian cells \cite{crissman1973rapid,kubitschek1983buoyant,basan2015inflating}, that the total volume is proportional to the total amount of proteins within the cell. Indeed, within these models, fixing the global concentration would immediately imply that the concentration of \textit{every} protein within the cell would be maintained  -- since as mentioned above the competition for the ribosomes naturally leads to the control of the \textit{relative} concentrations. How the global concentration is maintained is an important problem that is the subject of current research, both theoretically and experimentally, as we briefly review.

First, to appreciate the problem at hand, let us consider a model where cell envelope components (membranes, and in some microbes, the cell wall) are produced constitutively by the ribosomes -- without any feedback on the cellular density. How should we expect the density to behave within this model?

The answer depends on the cell geometry. Let us consider first the example of rod-shaped cells such as \textit{E. coli}. The continuous growth model of chapter \ref{chapter-contgrowth} led to exponential growth of the ribosome numbers. Under our assumption that a finite fraction of ribosomes are devoted to producing cell envelope components, we will find that, ignoring fluctuations, the cell \textit{surface area} should also increase exponentially. This conclusion would also be true when considering multiple generations and, correspondingly, the total surface area of the entire progeny of a given cell. Since both total mass and surface area grow exponentially with the same rate, and assuming that the cell geometry is fixed, density is expected to be confined to a relatively narrow range even within this simplified model. 

In fact, experimental data from various bacteria supports variants of this model: in such experiments, cell volume (rather than the ribosomal content) and surface area are concurrently quantified in bacterial cells during exponential growth. Ref. \cite{theriot} used data of this type to suggest a model where:
\be \frac{dS}{dt} \propto V. \label{theriot_eq}\ee
This phenomenological model predicts that when perturbing the surface/volume ratio, $S/V$, it will relax exponentially to its steady-state value, with a relaxation rate equal to the growth rate -- consistent with the experimental results for both \textit{E. coli} and \textit{C. crescentus}. Ref. \cite{shi2021precise} proposed a similar picture, albeit with a time-delay between surface and volume production. 

Recently, Ref. \cite{sven} suggested that, in fact, the dynamics of surface area growth should be described by:
\be \frac{dS}{dt} \propto \frac{dM}{dt},\label{sven_eq} \ee
with $M$ the cell mass. For exponential growth of volume and mass, Eqs. (\ref{theriot_eq}) and (\ref{sven_eq}) would be identical, but the two models make distinct predictions for time-varying growth conditions, and the experiments of Ref. \cite{sven} on \textit{E. coli} agreed better with Eq. (\ref{sven_eq}). A similar picture was also suggested to hold for \textit{B. subtilis} \cite{kitahara2022role}. 
Note that for rod-shape cells with a well-defined diameter, the constancy of the surface/mass ratio (up to small deviations, due to the change in the volume/surface area ratio during cell growth) is equivalent to a constant cellular density, since we may write:
\be \rho = M/V = (M/S) / (V/S), \ee
with the quantity $V/S$ governed solely by the cell's geometry, and Eq. (\ref{sven_eq}) leading to a constant $S/M$ ratio.
The same arguments applied to a \textit{spherical} cell would also lead to density homeostasis, albeit with non-negligible density fluctuations throughout the cell cycle, since the $S/V$ ratio changes more significantly in this case compared to that of rod-shaped cells. We note that currently there is no  experimental evidence to suggest these scaling laws, relating surface-area, volume, and mass, are applicable in mammalian cells. This could be due to the experimental challenges involved, as the complex and changing shapes of mammalian cells make quantification of surface area extremely difficult. 

One may view all of the aforementioned models coupling surface, volume, and mass growth as ``open circuits" -- the cells do not ``measure" concentration, and there is no direct feedback on it. However, effects such as molecular crowding may alter the diffusion of molecules and hence the rates of chemical reactions in the cell, providing precisely such feedback. For instance, Ref. \cite{alric2022macromolecular} demonstrated the effects of crowding on growth rate in budding yeast. However, the experiments were done using significant perturbations from the wild-type behavior, and whether crowding effects provide a biophysical regulatory cue for the small fluctuations around the typical density in organisms such as \textit{E. coli} remains to be seen.

A more subtle mechanism has recently been proposed to control cellular concentration \cite{rollin2023physical}. This work utilized the ``pump-and-leak" model, commonly used to describe the transport of ions through the cell membrane. In addition to highlighting the potential role of amino-acids in concentration homeostasis, their model explains the increase in volume and the concurrent decrease in cellular concentration upon entry to mitosis (the cell cycle phase where the chromosomes are segregated), known as ``mitotic swelling", observed in mammalian cells \cite{zlotek2015optical,son2015resonant,miettinen2022single}. An alternative mechanism could rely on the mechanical stress within the cell wall or membrane, as a means to ``measure" and respond to changes in the intra-cellular concentration \cite{mukherjee2023cell} (that would result in excesss osmotic pressure and hence mechanical stresses in the cell envelope). Such mechanics-based feedback is resonant with the theoretical and experimental results of Refs. \cite{amir2012dislocation, amir2014bending, wong2017mechanical}.

Another unresolved puzzle regards the relation between biomass growth, volume growth, and concentration homeostasis. In Chapter \ref{chapter-contgrowth} we discussed experimental evidence for the approximately exponential growth of biomass across multiple organisms, including the bacteria \textit{E. coli} and \textit{B. subtilis}. Recent analysis of single-cell microscopy data suggested that the \textit{volume growth} of these bacteria deviates from exponential growth, and is, in fact, faster than exponential (``super-exponential") \cite{nordholt2020biphasic,kar2021distinguishing}. If biomass growth is exponential but volume growth is not, cellular concentration is expected to be cell-cycle dependent -- yet, as discussed above, other measurements have suggested that, at least in \textit{E. coli}, it is not \cite{kubitschek1983buoyant}. These apparently contradictory observations are yet to be reconciled. Adding another wrinkle to the story, it was also  observed recently that the bacterium \textit{Mycobacterium tuberculosis} grows approximately \textit{linearly} at the single-cell level \cite{chung2023mycobacterium} -- hinting that, at least for this pathogen, the ribosome-centric framework of Chapter \ref{chapter-contgrowth} might not be adequate, as it would lead to exponential growth.

In mammalian cells, experiments measuring buoyant mass revealed cell-cycle-dependent deviations from exponential growth \cite{mu2020mass}, yet such deviations were not observed in microscopy measurements of cell volume \cite{cadart2018size} (though it should be noted that the cell types were not identical). 
Thus, here again, obtaining a consistent picture from mass and volume measurements remains to be achieved -- or, alternatively, establishing that cellular concentration is cell-cycle-dependent. Indeed, recent work in fission yeast suggested that cell density is not constant throughout the cell cycle \cite{odermatt2021variations}.

\subsection{Space: The final frontier}
\label{section=space}

A common feature of all models discussed in this Colloquium is that they do not explicitly consider the intracellular space: Ensemble means are described using ordinary differential equations in time, and fluctuations using the master equation, but nowhere do the spatial coordinates appear. Implicitly, the molecular encounters that underlie all cellular reactions are assumed to be diffusion driven, and to take place in a homogeneous, well-mixed environment. In the model equations, the rates of encounter are coarse-grained into the reaction rates of transcription, translation, degradation, etc. 

This approach, long the practice in modeling gene regulation \cite{paulsson2005models,bintu2005transcriptional}, is premised on the notion that the bacterial cell is sufficiently small, and internally uniform (lacking membranal compartments), such that the time it takes a molecule to diffuse across it ($< 1$ sec \cite{elowitz1999protein})\footnote{An intriguing caveat is that, in contrast to the normal (Fickian) diffusion of individual proteins, large molecular complexes have been shown to exhibit anomalous diffusion (specifically, sub-diffusion), with possible consequences for molecular encounter kinetics \cite{golding2006physical}.} is considerably shorter than the time scales for other processes under consideration, e.g., producing an mRNA or protein ($\sim$ minutes), replicating the genome or the cell ($\sim$ hour \cite{cooper1968chromosome,frederick1990physiology}). Under this premise, an explicit description of cellular spatiality is not required, and would unduly complicate models\footnote{A similar argument is commonly applied to eukaryotic cells, with the exception that the membrane-separated nucleus and cytoplasm are considered as different compartments, with molecules moving between them \cite{hansen2018cytoplasmic}. The oversimplification of this depiction is demonstrated by the recent report of concentration gradients inside yeast cells, gradients which could lead to intracellular differences in molecular diffusivity and reaction rates \cite{odermatt2021variations}.}.

However, there are reasons to suspect that the approximation of a homogeneous, diffusion-driven cell may be a poor one, even for bacteria. Bacterial cells have been revealed, in recent decades, to exhibit elements of spatial heterogeneity and organization, including, critically, in the machinery of gene expression. Considering {\it E. coli}, we now know that the key players in the gene expression process are localized preferentially to different regions of the cell, with only partial overlap between them: RNAP is enriched in the chromosomal region (the nucleoid), ribosomes are largely found at the periphery of the nucleoid, and the protein complex that degrades mRNA (the degradosome) is anchored to the cell membrane \cite{campos2013cellular}. Thus, the steps of gene expression conceivably proceed with spatial directionality, but this feature has only begun to receive theoretical attention \cite{castellana2016spatial}.

Beyond this regional preference, it was found that, as in eukaryotes \cite{cho2018mediator}, bacterial RNAP further exhibits areas of high localized concentration (reported to form through liquid/liquid phase separation \cite{ladouceur2020clusters}), which colocalize with genomic loci that encode ribosomes \cite{weng2019spatial,fan2023rna}. It is plausible that these RNAP clusters represent ``transcription factories" \cite{cook2010model}, regions of increased transcriptional activity, functioning to ensure sufficient production of ribosomes in the cell. If true, this would provide ribosomal genes with competitive advantage over other genes in securing the limiting resource of RNAP, under the scenario discussed in Section \ref{section-rnap-comp} above. 

Similarly, although the evidence in this regard is more limited, several studies suggest the local spatial accumulation of bacterial transcription factors (the proteins that modulate expression level), both at their genomic site of production \cite{kuhlman2013dna} and of targeted binding \cite{sarkar2018}. The experimental observations (again mirroring analogous ones in eukaryotic cells \cite{mir2017dense}), have been accompanied by several theoretical efforts to elucidate how spatial gradients of transcription factors form, and what the consequences are for their regulatory activity \cite{kuhlman2013dna,kolesov2007gene}.

The examples discussed above highlight the limitations of the spaceless picture, where models of gene expression still, for the most part, exist. The inadequacy of this modeling approach is further suggested by the fact that, whereas current models can typically reproduce the ensemble-averaged expression, they fail to correctly predict some aspects of the associated fluctuations, in particular, the near-universal scaling relation between the mean and noise, observed across different genes and even different organisms \cite{so2011general,sanchez2013genetic}. It may very well be that these failures reflect models' ignorance of intracellular space. Incorporating this feature thus remain a promising, and critical, direction for future research.

\section*{Acknowledgments}
We are grateful to Tianyou Yao for preparing the manuscript figures. We thank Andrew Pountain, Seunghyeon Kim, Sangjin Kim and Itai Yanai for sharing unpublished results. We are grateful to Sven van Teeffelen, Jie Lin and Teemu Miettinen for valuable comments. Work in the Golding lab is supported by the National Institutes of Health grant R35 GM140709 and by the Alfred P. Sloan Foundation. AA was supported by NSF CAREER 1752024 and by the Clore Center for Biological Physics.


%

\end{document}